%% file: manuscript.tex
\documentclass[twocolumn]{aastex63}
\usepackage{newtxtext,newtxmath}
\usepackage[T1]{fontenc}
\usepackage{ae,aecompl}
\usepackage{color}

\usepackage{graphicx}	
\usepackage{amsmath}	
\usepackage{amssymb}	
\usepackage{mathrsfs}
\usepackage{enumerate}
\usepackage{hyperref}

\received{XXX XX, XXXX}
\revised{XXX XX, XXXX}
\accepted{XXXX}
\submitjournal{ApJ}

\shorttitle{Dark matter profiles in the dSphs}
\shortauthors{Hayashi et al.}

\begin{document}

\title{Diversity of dark matter density profiles in the Galactic dwarf spheroidal satellites}

\correspondingauthor{Kohei Hayashi}
\email{k.hayasi@astr.tohoku.ac.jp}

\author[0000-0002-8758-8139]{Kohei Hayashi}
\affiliation{Astronomical Institute, Tohoku University \\
Aoba-ku, Sendai 980-8578, Japan}
\affiliation{Institute for Cosmic Ray Research, The University of Tokyo \\
Chiba 277-8582, Japan}

\author[0000-0002-9053-860X]{Masashi Chiba}
\affiliation{Astronomical Institute, Tohoku University \\
Aoba-ku, Sendai 980-8578, Japan}

\author[0000-0002-5316-9171]{Tomoaki Ishiyama}
\affiliation{Institute of Management and Information Technologies, Chiba University \\
1-33, Yayoi-cho, Inage-ku, Chiba 263-8522, Japan}

\begin{abstract}

The core-cusp problem is one of the controversial issues in the standard paradigm of $\Lambda$ cold dark matter~($\Lambda$CDM) theory. However, under the assumption of conventional spherical symmetry, the strong degeneracy among model parameters makes it unclear whether dwarf spheroidal (dSph) galaxies indeed have cored dark matter density profiles at the centers.
In this work, we revisit this problem using non-spherical mass models, which have the advantage of being able to alleviate the degeneracy.
Applying our mass models to the currently available kinematic data of the eight classical dSphs, we find that within finite uncertainties, most of these dSphs favor cusped central profiles rather than cored ones.
In particular, Draco has a cusped dark matter halo with high probability even considering a prior bias.
We also find the diversity of the inner slopes in their dark matter halos.
To clarify the origin of this diversity, we investigate the relation between the inner dark matter density slope and stellar-to-halo mass ratio for the sample dSphs and find this relation is generally in agreement with the predictions from recent $\Lambda$CDM and hydrodynamical simulations.
We also find that the simulated subhalos have anti-correlation between the dark matter density at 150~pc and pericenter distance, which is consistent with the observed one.
We estimate their astrophysical factors for dark matter indirect searches and circular velocity profiles, associated with huge uncertainties. 
To more precisely estimate their dark matter profiles, wide-field spectroscopic surveys for the dSphs are essential.

\end{abstract}



\section{Introduction} \label{sec:intro}

It is well documented that the concordant $\Lambda$ cold dark matter~($\Lambda$CDM) theory gives a remarkable description of the cosmological and astrophysical observations on large spatial scales such as the cosmic microwave background radiation~\citep[e.g.,][]{2011ApJS..192...18K,2018arXiv180706209P}, and large-scale structure of galaxies~\citep[e.g.,][]{2006Natur.440.1137S,2006PhRvD..74l3507T,2014MNRAS.439.2515O}.
At galactic and sub-galactic scales however, this theory has several discrepancies between the simulation predictions and observational facts~\citep[][for a review]{2017ARA&A..55..343B}.

One of them is the so-called ``core-cusp'' problem.
Dark-matter-only simulations based on the $\Lambda$CDM model have predicted a universal dark matter density profile with a strong cusp at the center~\citep[e.g.,][]{1994Natur.370..629M,1996ApJ...462..563N,1997ApJ...490..493N,1997ApJ...477L...9F,2013ApJ...767..146I}.
By contrast, the observations of dwarf spheroidal (dSph) and low surface brightness galaxies seem to favor a cored central dark matter density~~\citep[e.g.,][]{1995ApJ...447L..25B,2001MNRAS.323..285B,2007ApJ...663..948G,2008AJ....136.2761O,2010AdAst2010E...5D}.

To solve or ameliorate the issue, many possible solutions have been proposed. 
One of the solutions is to transform a cusped to cored central density through the baryonic effects such as stellar winds and supernova feedback~\citep[e.g.,][]{1996MNRAS.283L..72N,2002MNRAS.333..299G,2014ApJ...789L..17M,2016MNRAS.459.2573R} or heating of dark matter due to interaction of gas clumps and dark matter via dynamical friction~\citep[e.g,][]{2001ApJ...560..636E,2011MNRAS.418.2527I,2015MNRAS.446.1820N,DelPopolo:2015nda}.
Moreover, for the former mechanism, recent advanced simulations have predicted that the effect of core creation depends upon stellar mass and star formation history~\citep{2012MNRAS.422.1231G,2014MNRAS.441.2986D,2014MNRAS.437..415D,2015MNRAS.454.2092O,2016MNRAS.456.3542T,2017MNRAS.471.3547F,2018MNRAS.480..800H}.
Note that core formation ability of baryonic feedback is sensitive to the gas density threshold for a star formation, $n_{\rm sf}$, assumed in simulations.
A low threshold ($n_{\rm sf}=0.1$~cm$^{-3}$) is incapable of creating a core, while a high threshold ($n_{\rm sf}=10$-$1000$~cm$^{-3}$) is able to lead a core~\citep[e.g.,][]{2010Natur.463..203G,2019MNRAS.486.4790B}.
Although a high threshold is about four orders of magnitude greater than a low one, a whole range of the threshold is acceptable because current understanding of subgrid physics is not complete yet.

Another solution is, more radically, to replace CDM with other dark matter models that are well motivated from particle physics, such as~self-interacting dark matter~\citep[e.g.,][see also \citealt{2014arXiv1402.5143H,2015PhRvL.115b1301H}]{1992ApJ...398...43C,2000PhRvL..84.3760S,2016PhRvL.116d1302K,2018PhR...730....1T,Nadler:2020ulu}, and ultra-light dark matter~\citep[e.g.,][]{2000PhRvL..85.1158H,2014MNRAS.437.2652M,2014NatPh..10..496S,2016PhR...643....1M,2016PhRvD..94d3513S,2017MNRAS.471.4559M,2017PhRvD..95d3541H}.
These dark matter models can create a cored, low-dense central dark matter density profile on less massive-galaxy scales without relying on any baryonic physics.

Meanwhile, current dynamical studies for dSphs are challenged in the measurement of their central density profiles, because of the existence of $\rho_{\rm DM}-\beta_{\rm ani}$ degeneracy, where $\rho_{\rm DM}$ is a dark matter density and $\beta_{\rm ani}$ is a velocity anisotropy of stars as an unknown parameter~\citep[e.g.,][]{1982MNRAS.200..361B,1990AJ.....99.1548M,2009MNRAS.393L..50E}.
This degeneracy originates from the assumption that both stars and dark matter are spherically distributed and from the fact that only line-of-sight velocity components of stars are available from observations~\citep[e.g.,][]{2007ApJ...657L...1S}.
To disentangle this degeneracy, many dynamical modelings have been proposed, as exemplified by using higher order velocity moments~\citep[e.g.,][]{2002MNRAS.333..697L,2009MNRAS.394L.102L,1990AJ.....99.1548M}, virial theorem~\citep[e.g.,][]{2014MNRAS.441.1584R}, modeling multiple stellar populations~\citep[e.g.,][]{2008ApJ...681L..13B,2011ApJ...742...20W}, orbit-based dynamical models~\citep[e.g.,][]{2013ApJ...763...91J,2013MNRAS.433.3173B}, measuring the internal proper motion data~\citep{2018NatAs...2..156M,2019arXiv190404037M,2018ApJ...860...56S}, and non-parametric analysis~\citep[e.g.,][]{2017MNRAS.471.4541R,2019MNRAS.484.1401R}.
However, the inferred dark matter density profiles are not completely unified, and some of these models cannot distinguish a cusp from a core from the currently available kinematic data, due to considerable uncertainties in the derived dark matter density profiles and a prior bias of a dark matter inner slope parameter.
Thus, whether the central dark matter densities in dSphs are cored or cusped is yet unclear.

We emphasize that many of these studies assume spherical symmetry for both the stellar and dark components, even though we know both from observational facts and theoretical predictions that these components are actually non-spherical~\citep[e.g.,][]{2012AJ....144....4M,2018ApJ...860...66M,2006MNRAS.367.1781A,2014MNRAS.439.2863V}. 
In this paper, we relax the spherically symmetric assumption and perform the axisymmetric Jeans analysis for the dSphs.
Such non-spherical mass models have several advantages that (i)~giving the specific form of the distribution function is not required; (ii)~this analysis can treat {\it two-dimensional} distributions of line-of-sight velocity dispersions~\citep[e.g.,][]{2012ApJ...755..145H}, whereas it is impossible for spherical mass models; and (iii)~$\rho_{\rm DM}-\beta_{\rm ani}$ degeneracy can be mitigated~\citep{2008MNRAS.390...71C,BATTAGLIA201352,2015ApJ...810...22H}.
Several studies have developed axisymmetric mass models based on the Schwarzschild method~\citep{2012ApJ...746...89J} and Jeans anisotropic multiple Gaussian expansion model~\citep{2016MNRAS.463.1117Z}, but many of these assumed that a dark matter halo is still spherical while a stellar system is non-spherical.

Our group constructed, as presented in \citet{2015ApJ...810...22H}, totally axisymmetric dynamical mass models based on axisymmetric Jeans equations and applied the models to the dSphs with Milky Way and Andromeda galaxies~\citep[see also ][]{2012ApJ...755..145H}.
\citet{2016MNRAS.461.2914H} applied the axisymmetric mass models to the recent kinematic data for the ultra-faint dSphs as well as classical ones to evaluate the astrophysical factors for dark matter annihilation and decay with considering the uncertainties of non-sphericity.

Our previous models were yet incomplete in the point that an outer dark matter profile is fixed as $\rho_{\rm DM}\propto r^{-3}$ for the sake of simplicity.
Here, to step further from these previous studies, we adopt a generalized Herquist profile to explore a much wider range of physically plausible dark matter profiles and apply these non-spherical models to the latest observational data of the Galactic classical dSphs~(Draco, Ursa~Minor, Carina, Sextans, Leo~I, Leo~II, Sculptor, and Fornax) having a large number of member stars with well-measured radial velocities.

The paper is organized as follows. 
In Section 2, we explain axisymmetric models based on an axisymmetric Jeans analysis and our fitting procedure. 
In Section 3, we describe the photometric and spectroscopic data for the classical dSphs. 
In Section 4, we present the results of the fitting analysis. We also show the estimated dark matter density profiles and the values of astrophysical factors. 
In Section 5, we discuss the results of our estimations. 
Finally, conclusions are presented in Section 6.

\input{table1.tex}
\section{Models and Jeans analysis} \label{sec:jeans}
In this section, we briefly introduce our dynamical mass models in this work.
To show how precisely we are able to recover actual dark matter density profiles from our fitting analysis, we apply our mass models to mock data sets. The details about mock data and the results of mock analysis are shown in Appendix~\ref{sec:AppA}.

\subsection{Axisymmetric Jeans equations}
Assuming that a galaxy is in a dynamical equilibrium and collisionless under a smooth gravitational potential, the dynamics of stars in such a system is described by its phase-space distribution function governed by the steady-state collisionless Boltzmann equation~\citep{2008gady.book.....B}.
However, it is virtually impossible to solve this equation from the currently available data of stars in the dSphs whose positions along the line of sight are difficult to resolve and accurate proper motions are yet to be measured.
In order to alleviate this issue, one of the classical and useful approaches is to take moments of the equation.
The equations taking moments of the steady-state collisionless Boltzmann equation are the so-called Jeans equations.

For an axisymmetric and steady state system, the Jeans equations are expressed as 
\begin{eqnarray}
\overline{u^2_z} &=&  \frac{1}{\nu(R,z)}\int^{\infty}_z \nu\frac{\partial \Phi}{\partial z}dz,
\label{AGEb03}\\
\overline{u^2_{\phi}} &=& \frac{1}{1-\beta_z} \Biggl[ \overline{u^2_z} + \frac{R}{\nu}\frac{\partial(\nu\overline{u^2_z})}{\partial R} \Biggr] +
R \frac{\partial \Phi}{\partial R},
\label{AGEb04}
\end{eqnarray}
where $\nu$ is the three-dimensional stellar density and $\Phi$ is the gravitational potential, which is significantly dominated by dark matter for the Galactic dSphs.
The latter means that stellar motions in a system are governed only by a dark matter potential.
We assume that the cross terms of velocity moments such as $\overline{u_Ru_z}$ vanish and the velocity ellipsoid constituted by $(\overline{u^2_R},\overline{u^2_{\phi}},\overline{u^2_z})$ is aligned with the cylindrical coordinate.
We also assume that the density of tracer stars has the same orientation and symmetry as that of a dark halo.
$\beta_z=1-\overline{u^2_z}/\overline{u^2_R}$ is a velocity anisotropy parameter introduced by~\citet{2008MNRAS.390...71C}.
In this work, $\beta_z$ is assumed to be constant for the sake of simplicity\footnote{Nevertheless, this assumption is roughly in good agreement with dark matter simulations reported by~\citet{2014MNRAS.439.2863V} who have shown that simulated subhalos have an almost constant $\beta_z$ or a weak trend as a function of radius along each axial direction.}.
In principle, these second velocity moments are defined as $\overline{u^2}= \sigma^2+\overline{u}^2$, where $\sigma$ and $\overline{u}$ are dispersion and streaming motions of stars, respectively.
The latter streaming motions are small in the dSphs~\citep[e.g.,][]{2008ApJ...688L..75W}, and thus these galaxies are largely dispersion-supported stellar systems~\citep[e.g.,][]{2017MNRAS.465.2420W}.

To compare with the observed second velocity moments, the intrinsic second velocity moments derived by the Jeans equations are integrated along the line-of-sight second velocity moment followed by the previous works~\citep{1997MNRAS.287...35R,2006MNRAS.371.1269T,2012ApJ...755..145H}.
This moment can be written as 
\begin{equation}
\overline{u^2_{\rm l.o.s}}(x,y) = \frac{1}{I(x,y)}\int^{\infty}_{-\infty}\nu(R,z)\overline{u^2_{\ell}}(R,z)d\ell,
\label{los3}
\end{equation}
where $I(x,y)$ indicates the surface stellar density profile calculated from $\nu(R,z)$, and $(x,y)$ are the sky coordinates aligned with the major and minor axes, respectively.
$\overline{u^2_{\ell}}(R,z)$ is driven by
\begin{equation}
\overline{u^2_{\ell}} = \overline{u^2_{\ast}}\cos^2\theta + \overline{u^2_z}\sin^2\theta,
\label{los2}
\end{equation}
where $\theta$ is the angle between the line of sight and the galactic plane~($\theta=90^{\circ}-i$, which $i$ is an inclination angle explained below).
$\overline{u^2_{\ast}}$ is a velocity second moment derived from the projection $\overline{u^2_R}$ and $\overline{u^2_{\phi}}$ to the plane parallel with the galactic plane along the intrinsic major axis.
This moment is described as 
\begin{equation}
\overline{u^2_{\ast}} = \overline{u^2_{\phi}}\frac{x^2}{R^2} + \overline{u^2_R}\Bigl(1-\frac{x^2}{R^2}\Bigr).
\label{los1}
\end{equation}

\subsection{Stellar density profile}\label{sec:starprof}
For the stellar density profile, we adopt a Plummer profile \citep{1911MNRAS..71..460P} generalized to an axisymmetric shape:
\begin{eqnarray}
\nu(R,z)=\frac{3L}{4\pi b^3_{\ast}}\frac{1}{(1+m^2_{\ast}/b^2_{\ast})^{5/2}}
\label{plummer}
\end{eqnarray}
where $m^2_{\ast}=R^2+z^2/q^2$, so that $\nu$ is constant on spheroidal shells with an intrinsic axial ratio $q$, and $L$ and $b_{\ast}$ are the total luminosity and the half-light radius along the major axis, respectively.
This profile can be analytically derived from the surface density profile using Abel transformation: $I(x,y)=(L/\pi b^2_{\ast})(1+m^{\prime 2}_{\ast}/b^2_{\ast})^{-2}$,
where $m^{\prime 2}_{\ast}=x^2+y^2/q^{\prime 2}$.
$q^{\prime}$ is a projected axial ratio and is related to the intrinsic one $q$ through the inclination angle $i$~$(=90^{\circ}-\theta)$: $q^{\prime 2} = \cos^2i+q^2\sin^2i$.
This equation can be rewritten as $q=\sqrt{q^{\prime 2}-\cos^2i}/\sin i$, and thus the allowed range of the inclination angle is bounded with $0\leq\cos^2i<q^{\prime 2}$.
In this work, we assume that the stellar distribution has an {\it oblate} shape only.
This is motivated by the result from \citet{2015ApJ...810...22H} which concluded that most of stellar distributions of the dSphs are much better fitted by the oblate shape than by the prolate ones.

\subsection{Dark matter density profile}\label{sec:dmprof}
In this work, we adopt a generalized Hernquist profile given by~\citet{1990ApJ...356..359H} and also~\citet{1996MNRAS.278..488Z} with considering non-spherical dark matter halos,
\begin{eqnarray}
&& \rho_{\rm DM}(R,z) = \rho_0 \Bigl(\frac{r}{b_{\rm halo}} \Bigr)^{-\gamma}\Bigl[1+\Bigl(\frac{r}{b_{\rm halo}} \Bigr)^{\alpha}\Bigr]^{-\frac{\beta-\gamma}{\alpha}},
 \label{DMH} \\
&& r^2=R^2+z^2/Q^2,
\label{DMH2}
\end{eqnarray}
where $\rho_0$ and $b_{\rm halo}$ are the scale density and radius, respectively, $\alpha$ is
the sharpness parameter of the transition from the inner slope $\gamma$ to the outer slope $\beta$, 
and $Q$ is a constant axial ratio of a dark matter halo. 
This model covers a broad range of physically plausible dark matter profiles from the cusped Navarro-Frenk-White~\citep[hereafter NFW,][]{1997ApJ...490..493N} profile to the cored Burkert profile~\citep{1995ApJ...447L..25B}.

\input{table2.tex}

\subsection{Fitting procedure}\label{sec:fitting}
In order to estimate the dark matter density profiles in the dSphs, we explore the most likely parameter values by fitting theoretical and observed second velocity moments of each dSph.
In this work, we suppose that the line-of-sight velocity distribution is Gaussian and centered on the systemic velocity of the galaxy $\langle u \rangle$.
Given that the total number of member stars for each dSph is $N$, and the observed line-of-sight velocity of the $i$th member star and its velocity error is expressed by $u_i\pm\delta_{u,i}$ at the sky plane coordinates~$(x_i,y_i)$, the likelihood function is described as
\begin{equation}
{\cal L} = \prod^{N}_{i=1}\frac{1}{(2\pi)^{1/2}[(\delta_{u,i})^2 + (\sigma_i)^2]^{1/2}}\exp\Bigl[-\frac{1}{2}\frac{(u_i-\langle u \rangle)^2}{(\delta_{u,i})^2 + (\sigma_i)^2} \Bigr],
\label{LF}
\end{equation}
where $\sigma_i$ is the theoretical line-of-sight velocity dispersion at~$(x_i,y_i)$ which is calculated by model parameters and the Jeans equations.
The systemic velocity $\langle u \rangle$ of the dSph is a nuisance parameter that we marginalize over as a flat prior.
For the model parameters, we introduce flat or log-flat priors over the following ranges:
\begin{enumerate}[(i)]
\item $0.1\leq Q\leq2.0$;
\item $0.0\leq \log_{10}[b_{\rm halo}/{\rm pc}]\leq5.0$;
\item $-5.0\leq \log_{10}[\rho_0/(M_{\odot}~{\rm pc}^{-3})]\leq5.0$;
\item $-1.0\leq -\log_{10}[1-\beta_{z}]<1.0$;
\item $0.5\leq\alpha\leq3$
\item $3.0\leq\beta\leq10$
\item $0.0\leq\gamma\leq2.5$
\item $\cos^{-1}(q^{\prime})< i/{\rm deg} \leq 90.0$.
\end{enumerate}

In order to obtain the posterior probability distribution function (PDF) of each parameter by the above likelihood function, we perform a Markov Chain Monte Carlo~(MCMC) techniques, based on Bayesian parameter inference, using the Metropolis-Hastings algorithm~\citep{1953JChPh..21.1087M,10.1093/biomet/57.1.97}.
To avoid an influence of initial conditions and to generate independent samples, we take several post-processing steps such as burn-in step, the sampling step and length of the chain.
We evaluate the percentiles of these PDFs to estimate credible intervals for each parameter straightforwardly.

\subsection{Results from applying our models to mock data}
To scrutinize whether our models can reproduce dark matter density profiles, we apply them to mock data.
Here, we focus on testing how precisely our models are able to reproduce inner density slopes of dark halos.
In this subsection, we briefly describe the procedure for mock analysis and summarize the result from this analysis. The details are shown in Appendix~\ref{sec:AppA}.

We use public mock data sets provided by~\citet{2016MNRAS.463.1117Z}, which generated kinematic samples with axisymmetric stellar systems embedded in spherical dark matter halos and generated two kinds of dark matter halos: one with a cusped halo and one with a cored halo.

When we apply our models to these mock data, we first estimate a projected axial ratio and a half-light radius, employing a maximum likelihood analysis~\citep{2008ApJ...675..201M} with the Plummer stellar density profile.
Then, we perform a MCMC analysis for mock kinematic samples to estimate the dark matter density profiles. 

In this mock analysis, we carry out the fitting in the three cases: (A) a cusped model with 1000 samples, (B) a cored model with 1000 samples, and (C) a cored one with 4000 samples.
The dark matter density profiles estimated from the analysis are nearly reproduced within $1\sigma$ uncertainties for each case.
The estimated values of dark matter inner slope, $\gamma$, are $\gamma=1.2^{+0.4}_{-0.5}$, $0.4^{+0.8}_{-0.3}$, and $0.4^{+0.3}_{-0.2}$ for the cusped mock~(A), the cored ones~(B) and (C), respectively~(see Figure~\ref{fig:Mock}).
It is found that the MCMC fitting to the cored mock data results in a somewhat biased dark matter density profile with $\gamma \simeq 0.4$, although the model is nearly consistent with a cored halo within $1\sigma$ confidence.
Such a bias is also appeared in several previous works~\citep[e.g.,][]{2016MNRAS.463.1117Z,2018MNRAS.481..860R}, but the reason for this bias is yet unclear.

Consequently, we should bear in mind the fact that there exits this small amount of bias for a dark matter inner slope in our fitting analysis.

\section{Data}\label{sec:data}
In this section, we present the basic properties of photometric and spectroscopic data of the eight classical dSphs: Draco, Ursa~Minor, Carina, Sextans, Leo~I, Leo~II, Sculptor, and Fornax.
The classical dSphs have a larger number of line-of-sight velocities for the resolved stars ($\gtrsim200$~stars) than ultra faint dwarf galaxies.
These galaxies also have large velocity dispersions~($\gtrsim10$~km~s$^{-1}$), so that an influence of unresolved binary stars on the velocity dispersion measurements of each galaxy can be negligible~\citep{2010ApJ...721.1142M,2013ApJ...779..116M,2017ApJ...836..202S,2018AJ....156..257S}.

Table~\ref{table1} lists the observational properties of the eight dSphs:~the number of member stars with velocity measurements available from the kinematic analysis, the central sky coordinates, distances from the Sun, projected half-light radii, projected stellar axial ratios, systemic velocities, and their references.
Following previous works, we fix the values of distance, half-light radius, and axial ratio of dSphs in this paper.

For the stellar kinematic data of their member stars, we use the published data as follows.
For Carina, Draco, Ursa~Minor, Leo~I, and Leo~II, we use the stellar-kinematic data taken from \citet{2016ApJ...830..126F}, \citet{2015MNRAS.448.2717W}, \citet{2018AJ....156..257S}, \citet{2008ApJ...675..201M}, \citet{2017ApJ...836..202S}, respectively.
For Sextans, Sculptor, and Fornax, we use the data published by \citet{2009AJ....137.3100W,2009AJ....137.3109W}.
The membership selection criteria for each galaxy follow the methods described by each of the observational papers given above.

\section{Results} \label{sec:results}
In this section, we present the results from the MCMC fitting analysis described above and several trends among the resultant parameters.
Moreover, as a by-product of the fitting results, we estimate the astrophysical factors for dark matter annihilation and decay.

\subsection{Best-fitting models} \label{sec:best}
Table~\ref{table2} shows the best-fitting parameters for each dSph.
The error values indicate the 68~per~cent credible intervals computed from posterior PDFs of the parameters. 
We also show the dark matter density at 150~pc, $\rho_{\rm DM}(150\ {\rm pc})$, to compare with the other works. 
\citet{2019MNRAS.484.1401R} presented the dark matter density at a common radius of 150~pc from the center of each galaxy, $\rho_{DM}(150\ {\rm pc})$, which is insensitive to the choice of a $\gamma$'s prior in spherical mass models.
Furthermore, using this density, \citet{2019MNRAS.490..231K} pointed out the anti-correlation between $\rho_{DM}(150\ {\rm pc})$ and their orbital pericenter distances, $r_{\rm peri}$, of the classical dSphs.
This implies a survivor bias: while galaxies with low dark matter densities were completely destroyed by strong tidal effects, those with high dark matter densities survive in the present day.
Following these works, we also calculate the dark matter density at 150~pc along the major axes of the sample dSphs, considering non-sphericity of a dark matter halo, and the calculated $\rho_{DM}(150\ {\rm pc})$ are tabulated in the last column of Table~\ref{table2}.
We discuss it in the next section.

\begin{figure*}
	\begin{minipage}{0.49\hsize}
		\begin{center}
			\includegraphics[width=\columnwidth]{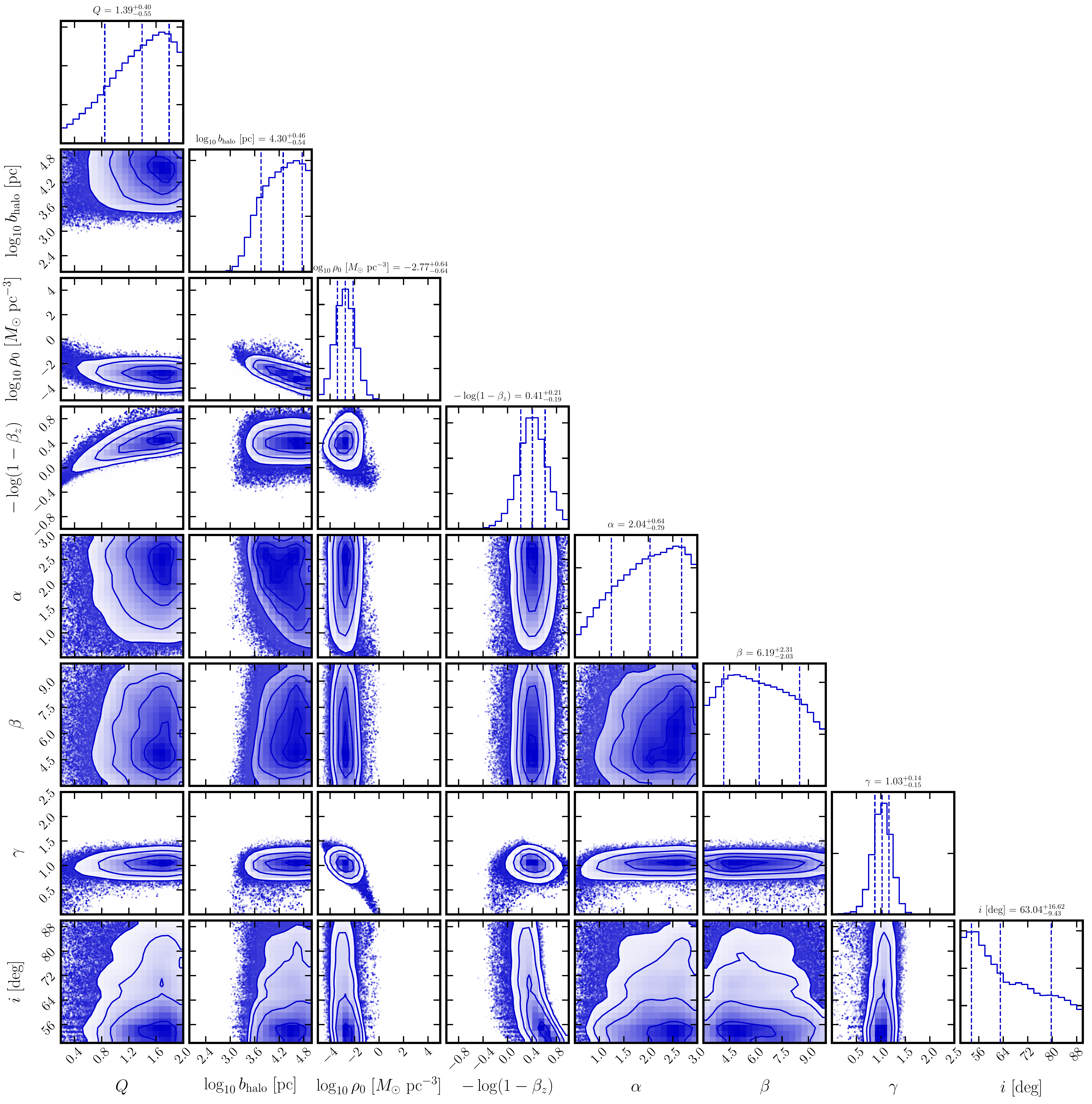}
		\end{center}
	\end{minipage}
	\begin{minipage}{0.49\hsize}
		\begin{center}
			\includegraphics[width=\columnwidth]{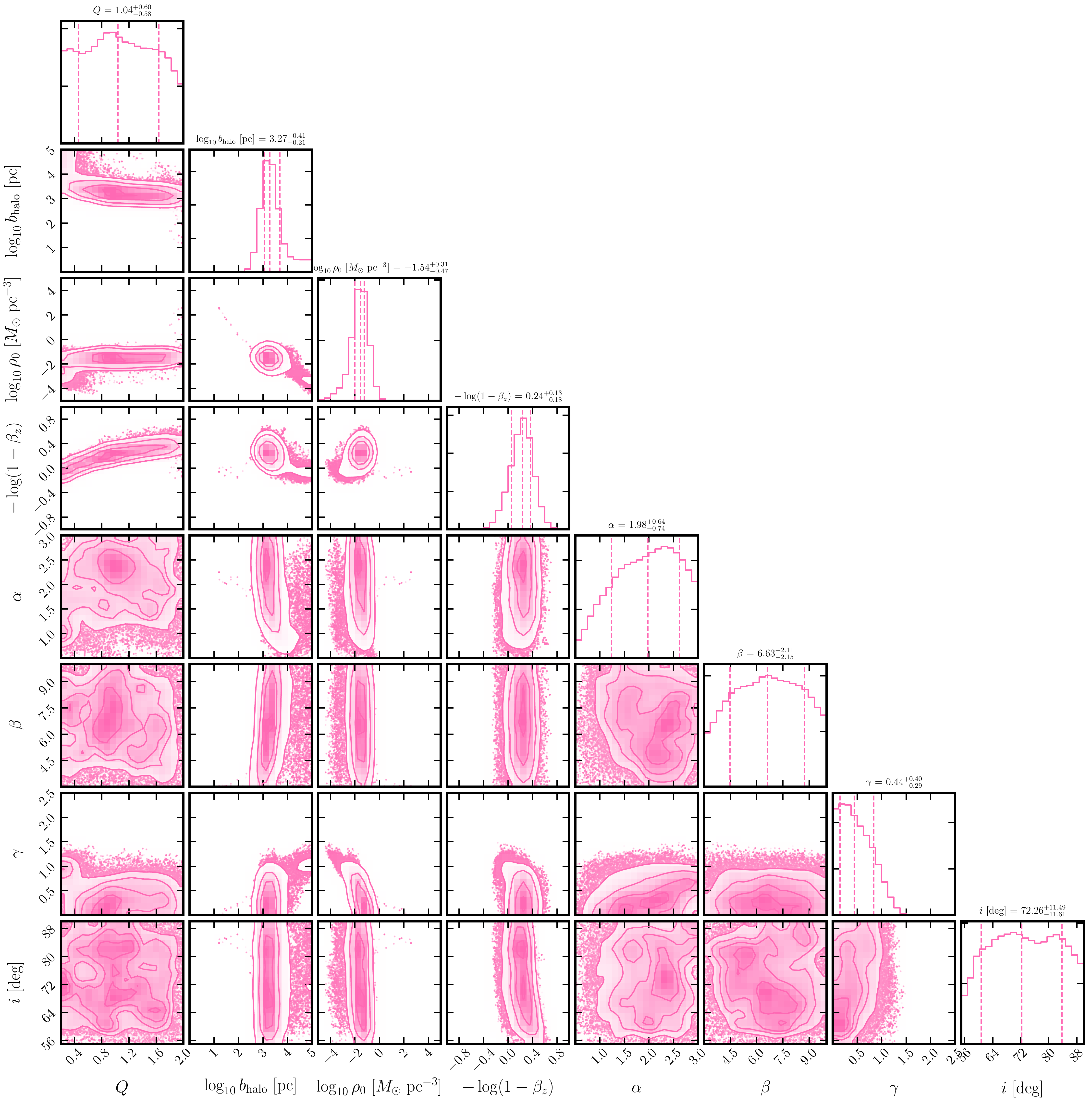}
		\end{center}
	\end{minipage}
    \caption{Posterior distributions for the fitting parameters for Draco~(left) and Fornax~(right).
    The dashed lines in each histogram represent the median and 68~per~cent confidence values. The contours in each panel are the 68, 95, and 99.7~per~cent regions.}
    \label{drafnx}
\end{figure*}

Figure~\ref{drafnx} displays the posterior PDFs implemented by the MCMC fitting for Draco~(left panel) and Fornax~(right panel) dSphs as the representative galaxies in our sample.
The results of other galaxies are similar to these and thus we show the other posterior PDFs~(Figure~\ref{carumi}, \ref{leo12}, and \ref{sclsex}) in Appendix~\ref{sec:AppB}. 
The contours in these figures show 68, 95, and 99.7 per~cent credible interval levels.
The vertical lines in each histogram also show the median and 68 per~cent credible interval levels.
From these posterior PDF maps, the parameters $Q$, $\alpha$, $\beta$, and $i$ are widely distributed in these parameter ranges and thus it is difficult to get limits on them.
On the other hand, the other parameters $b_{\rm halo}$, $\rho_0$, $\beta_z$, and $\gamma$ are better constrained than the above parameters, even though there are obvious degeneracies between $b_{\rm halo}$-$\rho_0$ and $Q$-$\beta_z$, which have already been discussed in several previous papers~\citep[e.g.,][]{2008MNRAS.390...71C,BATTAGLIA201352,2015ApJ...810...22H}.
In particular, the velocity anisotropy parameters $\beta_z$ of all galaxies (see also Table~\ref{table2}) tend to be a somewhat radially-biased velocity ellipsoid.
Moreover, it is worth noting that owing to non-spherical models, the inner slope parameter of a dark matter density profile $\gamma$ can be confined without being distracted by any parameter degeneracies.
From these PDFs and Table~\ref{table2}, we find that the posteriors of $\gamma$ show a wide spread from cusped $(\gamma>1.0)$ to shallower cusped $(\gamma<0.5)$ inner dark matter density slopes, even though there is a large uncertainty.
We discuss this further in Section~\ref{sec:bestdmprof}.

\begin{figure*}
\begin{center}
\includegraphics[scale=0.33]{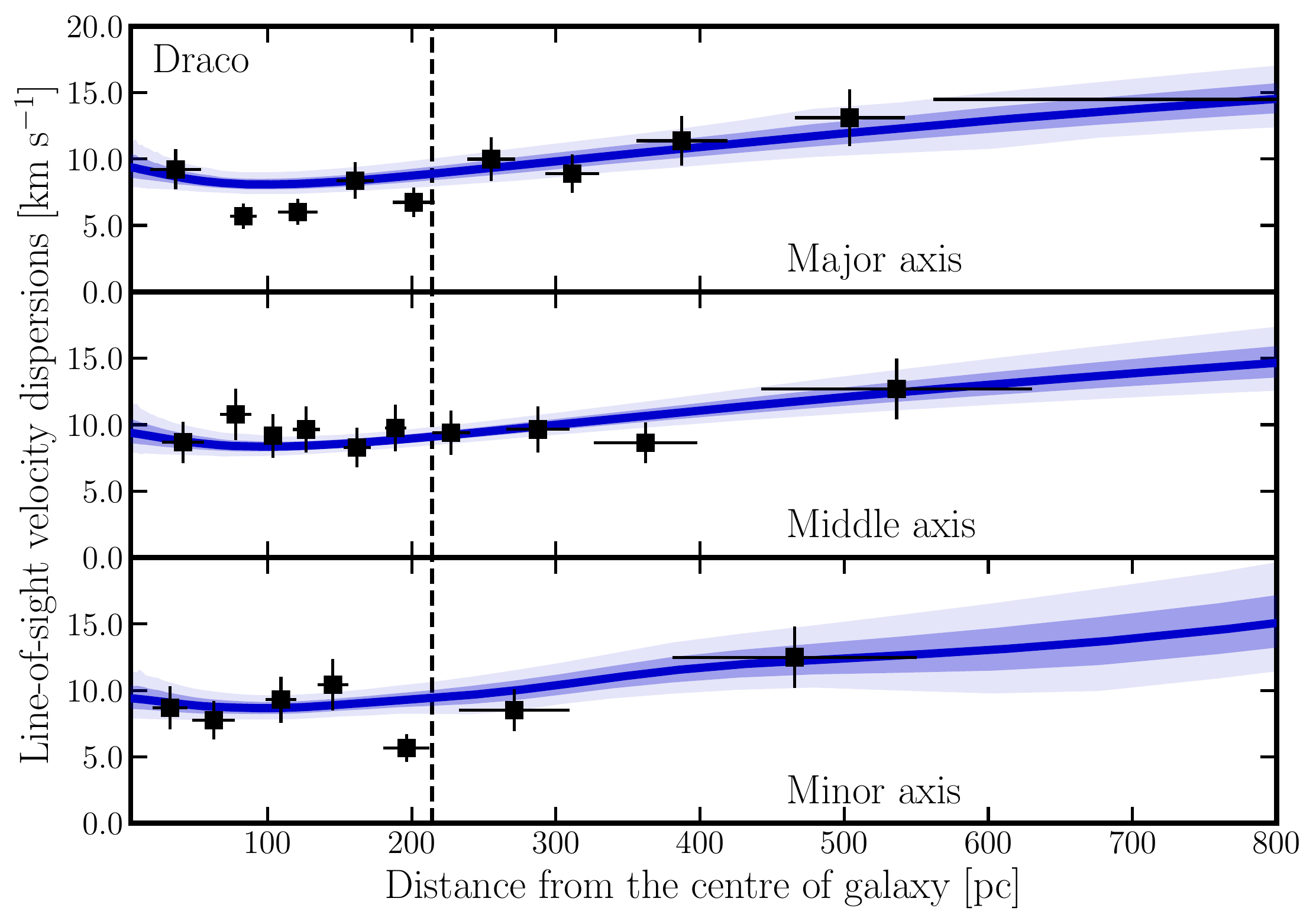}
\includegraphics[scale=0.33]{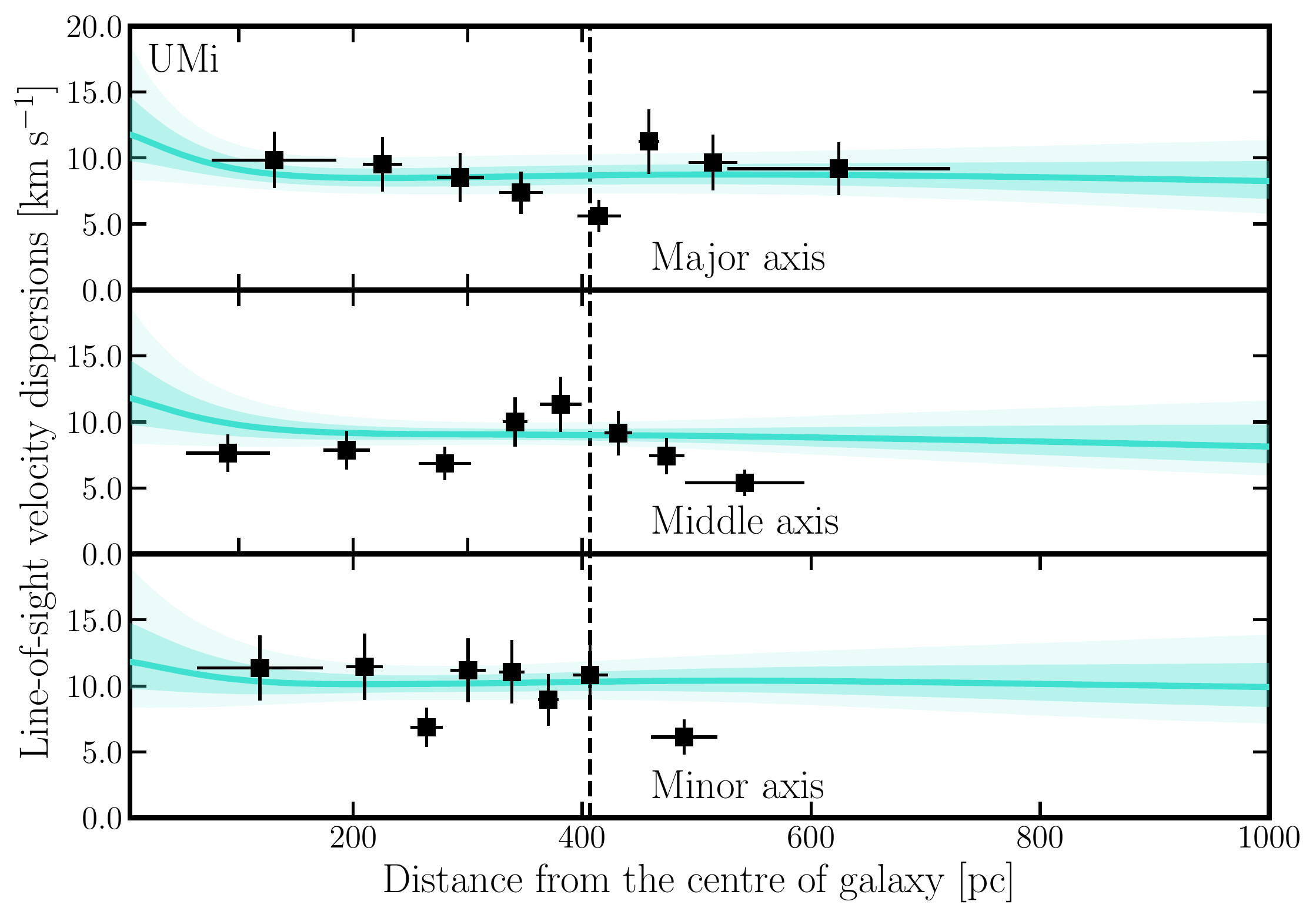}
\includegraphics[scale=0.33]{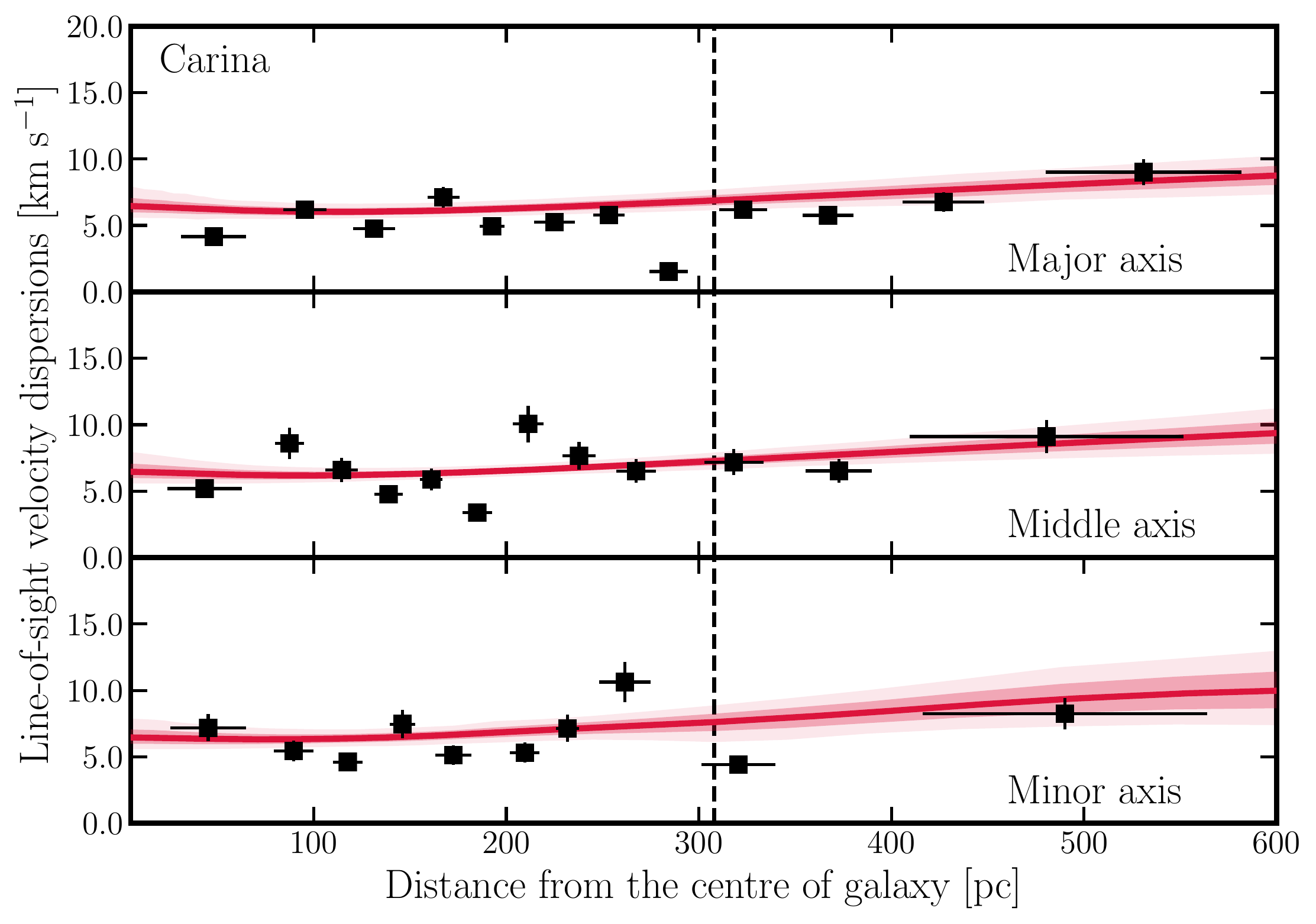}
\includegraphics[scale=0.33]{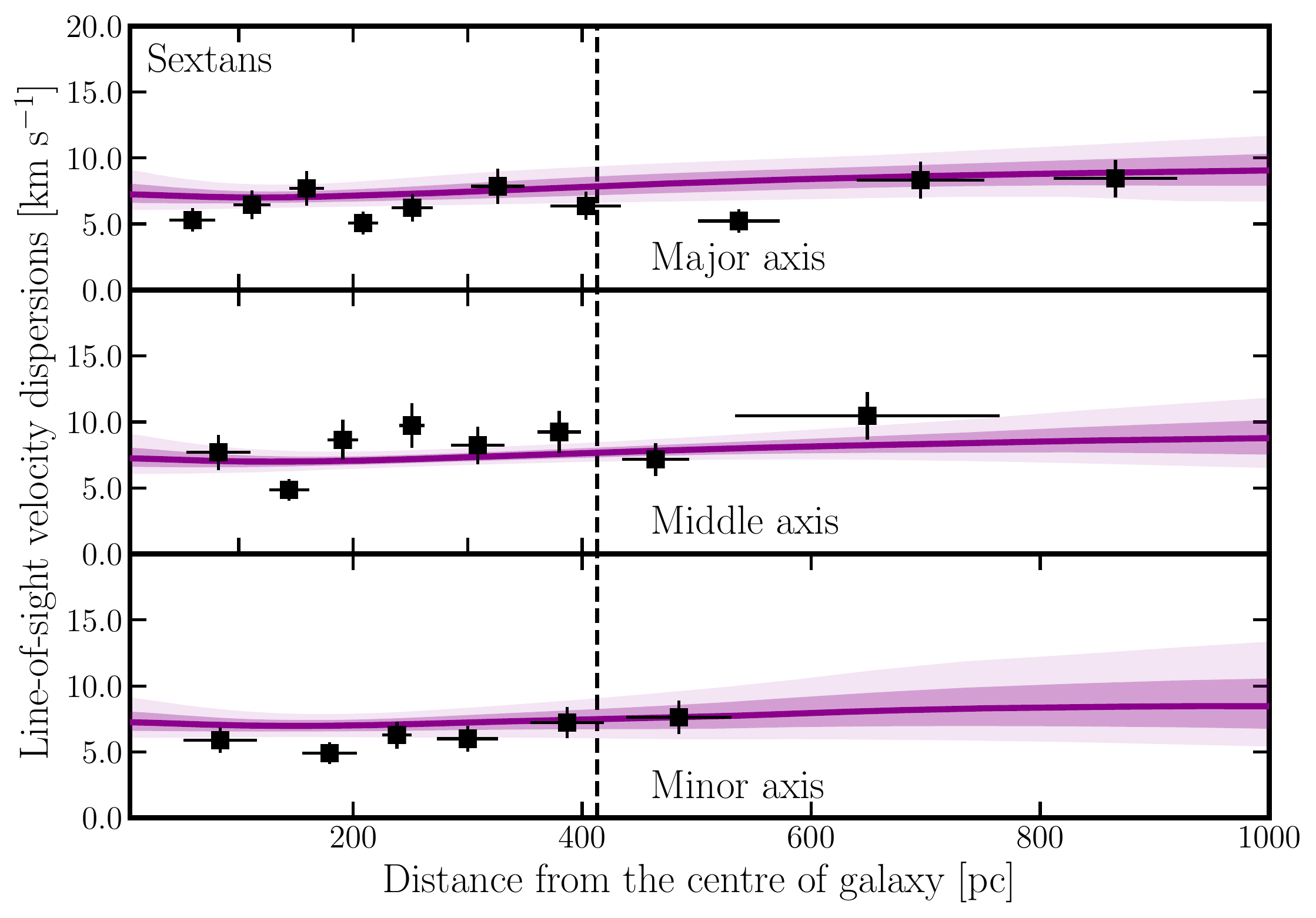}
\includegraphics[scale=0.33]{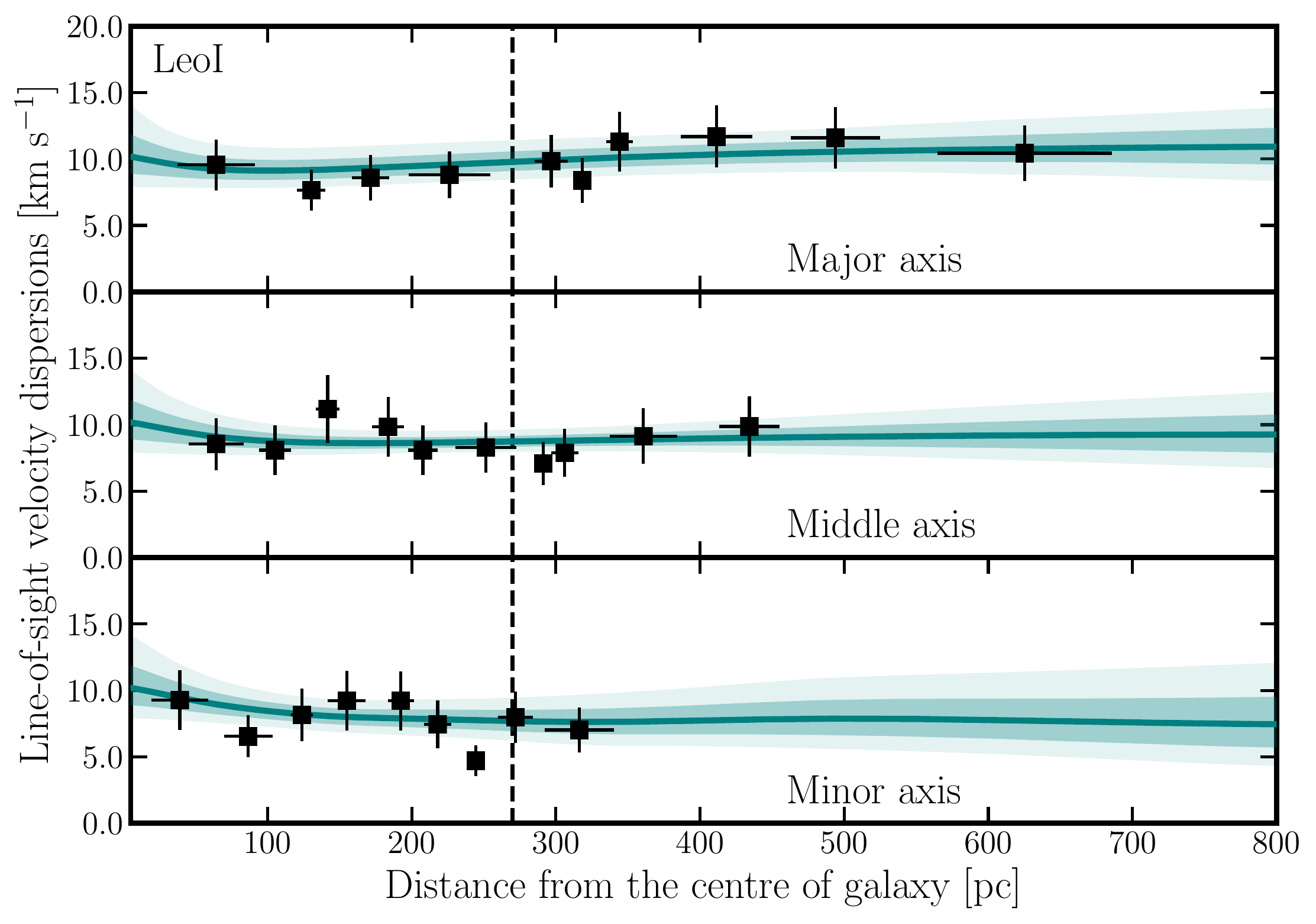}
\includegraphics[scale=0.33]{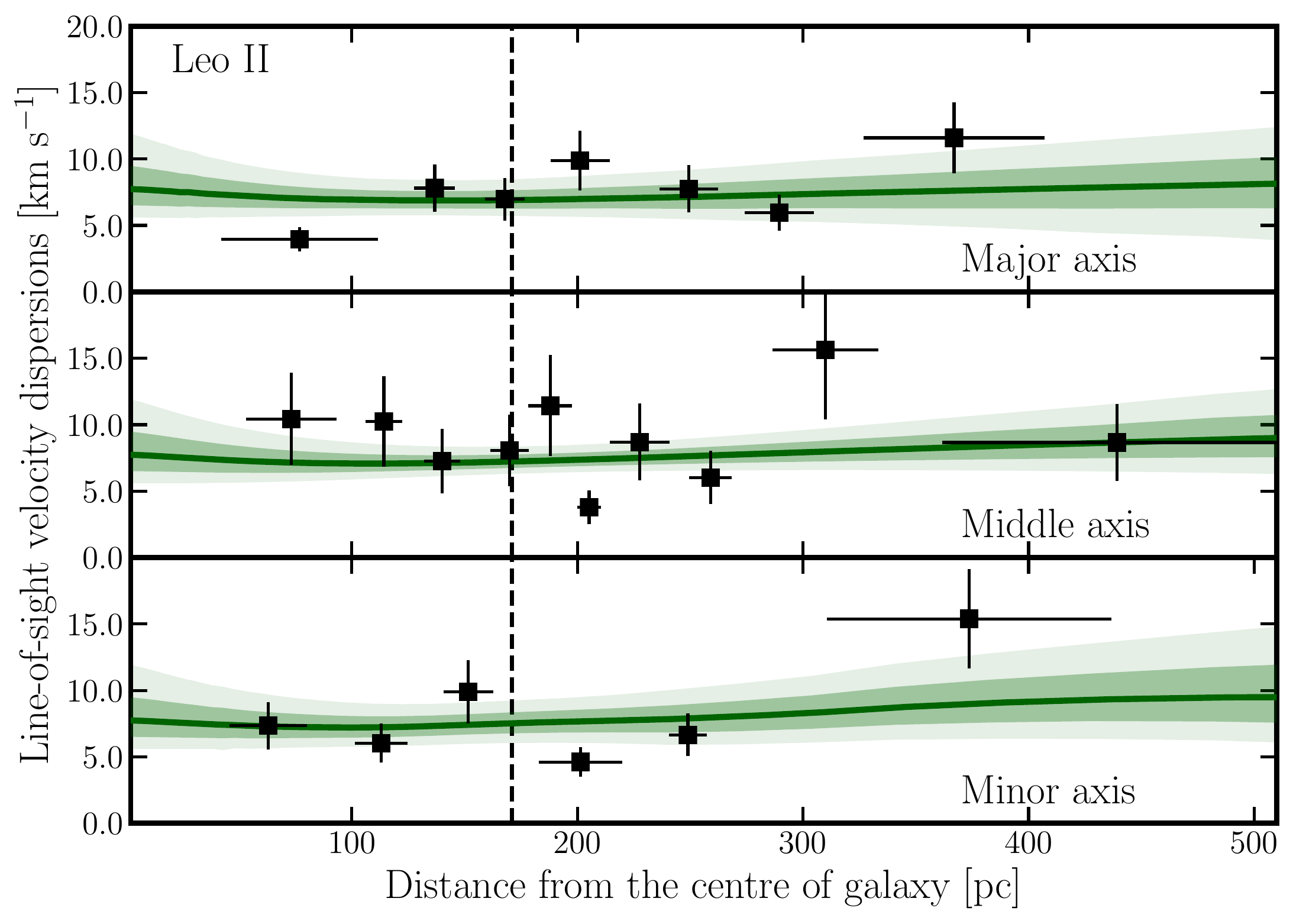}
\includegraphics[scale=0.33]{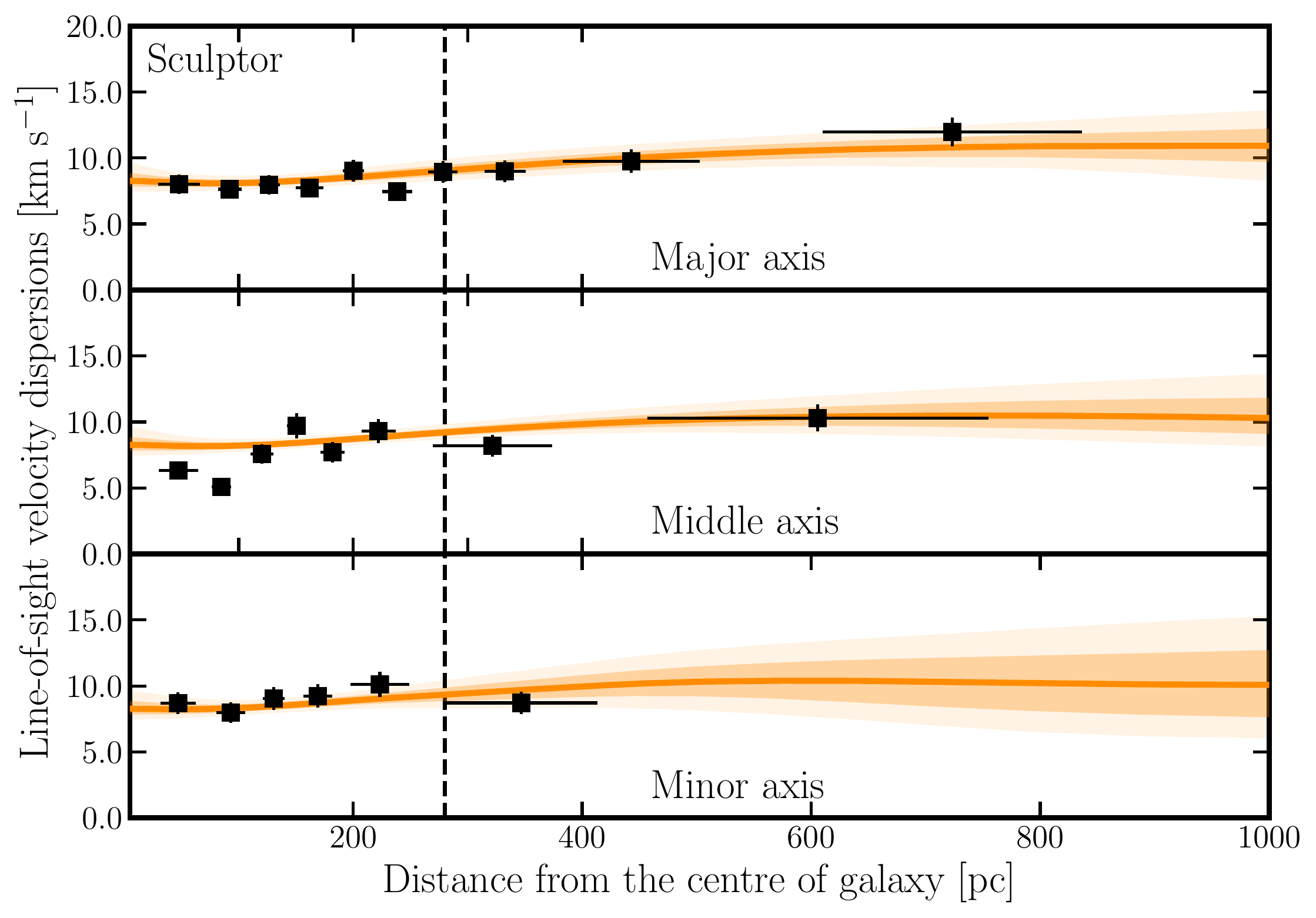}
\includegraphics[scale=0.33]{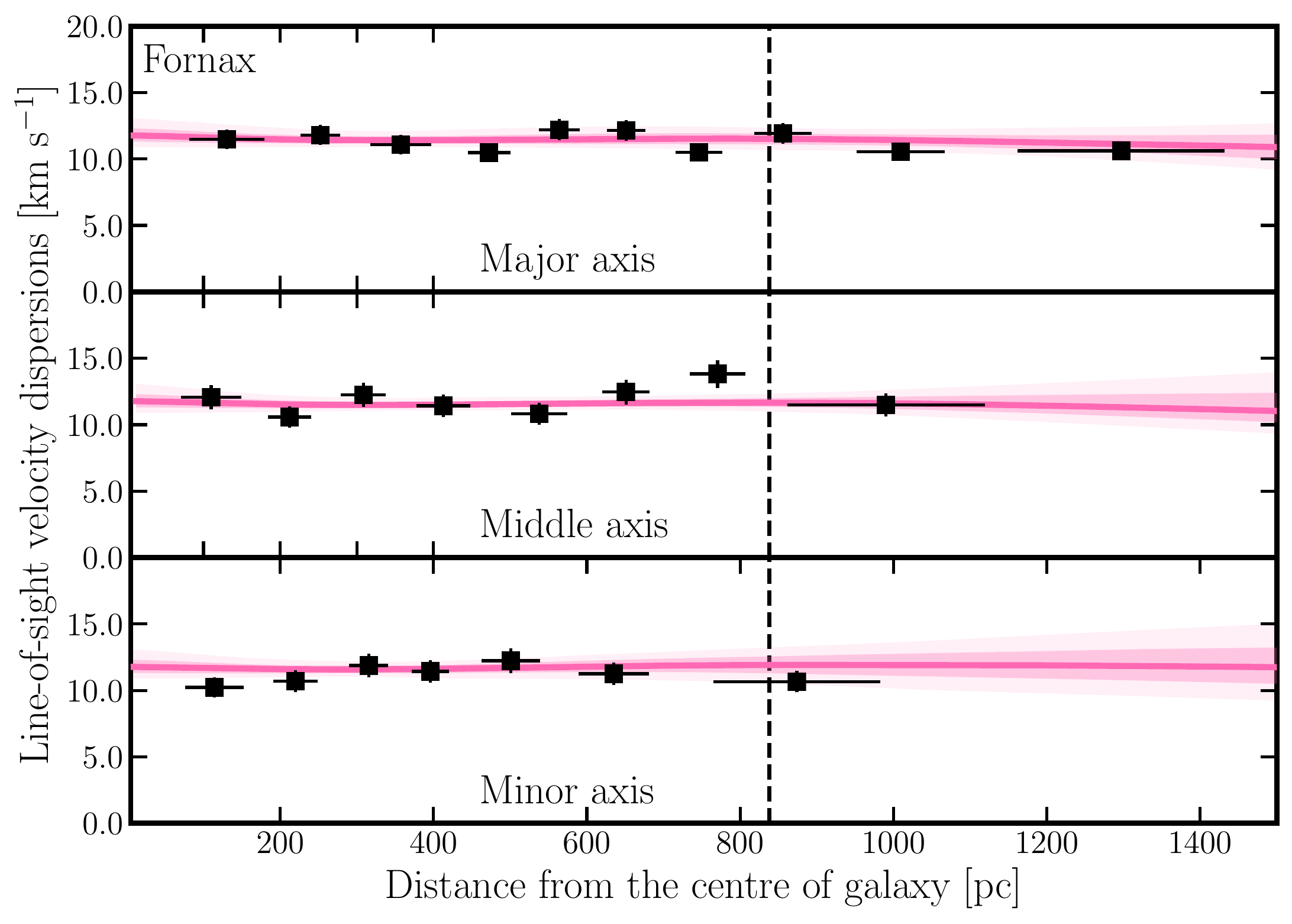}
\end{center}
\caption{Line-of-sight velocity dispersion along major, middle and minor axes for each dSph. The black squares with error bars in each panel denote the observed ones. The solid lines are the median velocity dispersion of the models and the dark and light shaded regions encompass the 68~per cent and 95~per cent confidence levels from the results of the unbinned MCMC analysis. The vertical dashed lines in each panel correspond to their half-light radii.}
\label{los}
\end{figure*}

Figure~\ref{los} shows the comparison between the observed and the estimated line-of-sight velocity dispersion profiles obtained by the resultant posterior PDFs, to present our fitting analysis successfully reproduced to the binned data\footnote{The method for calculating these binned profiles along the projected major, middle, and minor axes for the dSphs is the same way as \citet{2019MNRAS.tmp.2554H}, and thus the details are described in the Section~3.1.2 in that paper.}.
In this figure, the colored solid lines and shaded regions denote the median and confidence levels (dark: 68~per~cent, light: 95~per~cent) of our {\it unbinned} MCMC analysis.
The black points with error bars denote {\it binned} velocity dispersions calculated by the observed data.
These errors correspond to the 68~per~cent confidence intervals.
As shown in this figure, our mass models and unbinned analysis can provide good fits to the binned data for all dSphs.

\subsection{Revisiting the core-cusp problem} \label{sec:bestdmprof}
\subsubsection{Dark matter density profiles}
Using the results of the MCMC fitting analysis for the kinematic data of the dSphs, we estimate the dark matter density profiles by marginalizing all free parameters.
Figure~\ref{dmpro} shows the inferred dark matter density profiles of all sample dSphs.
The solid lines show the medium, and dark and light contours mark the 68~per~cent and 95~per~cent intervals.
The vertical black lines mark the projected half light radii of each dSph.

Firstly, it is noteworthy that in our non-spherical models, Draco favors a cusped inner slope for its dark matter density profile, which is consistent with an NFW cusp predicted by $\Lambda$CDM theory.
Even if we consider 95~per~cent confidence intervals of the dark matter profile, its inner slope still remains cuspy.
Therefore, Draco highly likely has a cusped dark matter halo.

Secondly, Ursa~Minor, Leo~I and Leo~II also prefer cusped dark matter halos, although the uncertainties for the inner slope, $\gamma$, are larger than for Draco.
On the other hand, the remaining sample of dSphs (Carina, Sextans, Sculptor, and Fornax) favors smaller $\gamma$ and thus has less dense than the other dSphs which have cuspy dark matter halos.
In particular, Sextans, Sculptor, and Fornax permit $\gamma=0$, i.e. a cored dark matter density within their 95~per~cent confidence intervals.

Notably, the dark matter density profile in Draco is better constrained than Fornax, though the data volume of Fornax is greater than Draco.
This is because while the observed kinematic sample in Draco covers the stars up to its outskirts, that in Fornax is limited only to its inner region.
Actually, \citet{2015ApJ...810...22H} suggested that the lack of kinematic sample volume in the outer region of a galaxy makes the constraints on the dark matter profile very uncertain.

Therefore, from our dynamical analysis, we propose that there is {\it no} core-cusp problem in the Galactic classical dSphs.
Moreover, a {\it diversity} of the inner density slope, $\gamma$, is found for these dSphs.
Note that this result is in agreement with \citet{2019MNRAS.484.1401R}, which investigated the inner dark matter densities in the Galactic dSphs as well as in low surface brightness galaxies based on non-parametric spherical Jeans analysis.

\subsubsection{Why do some galaxies prefer cusped dark matter halos?} \label{sec:demonstration}
As shown in the previous section, we present that some dSphs prefer cusped dark matter density profiles.
Then the question is why these galaxies are regarded to have cusped dark matter halos.
We schematically illustrate this reason in Figure~\ref{los_demo4} in the Appendix~\ref{sec:AppC}.
This figure shows the normalized line-of-sight velocity dispersion profiles along the major~(top panels) and the minor axes~(bottom panels) for the oblate stellar system $(q=0.7)$.
The left-hand panels show the dispersion profiles with changing the value of velocity anisotropy parameter, $\beta_z$, under spherical dark matter halo, $Q=1$, whilst the right-hand ones depict those with changing $Q$ under $\beta_z=0$.

As already discussed in \citet{2008MNRAS.390...71C} and \citet{2015ApJ...810...22H}, the variation of $Q$ and $\beta_z$ gives a similar effect on line-of-sight dispersion profiles.
For instance, as is shown in the top-left panel of Figure~\ref{los_demo4}, the effect of $\beta_z>0$ (i.e., red lines) increases inner line-of-sight velocity dispersions and decreases outer ones, simultaneously, compared with those in the fiducial $(Q=1,\beta_z=0)$ case which corresponds to the black lines.
In the top-right panel, the effect of $Q<1$ (the red lines) is resemblant in the features of line-of-sight velocity dispersion profiles computed by $\beta_z>0$ (the red lines in the top-left panel), even though there is a difference between these effects at the outer parts (the reason of this difference is already discussed in \citealt{2015ApJ...810...22H}). 

However, comparing the dispersion profiles in the cases for cusped (the solid lines) and for cored (the dotted lines) dark matter density profile, we can see a difference in the shape of those profiles at inner parts.
In the case of a cusped dark matter halo, the velocity dispersion profiles along both major and minor axes rapidly increase towards the central region, while there is no such trend in the case of a cored one.
Looking at the observed line-of-sight velocity dispersion profiles in Figure~\ref{los}, Draco, as an example, seems to have the trend characterized by a cusped dark matter halo, whereas Fornax has the almost flat profiles.
Therefore, we suggest that Draco highly likely has a cusped dark matter halo.
It is also found that the feature of a central velocity dispersion profile can be important in determining an inner slope of a dark matter density profile.

\subsubsection{The robustness of our results}
\begin{figure*}[t!]
	\includegraphics[scale=0.4]{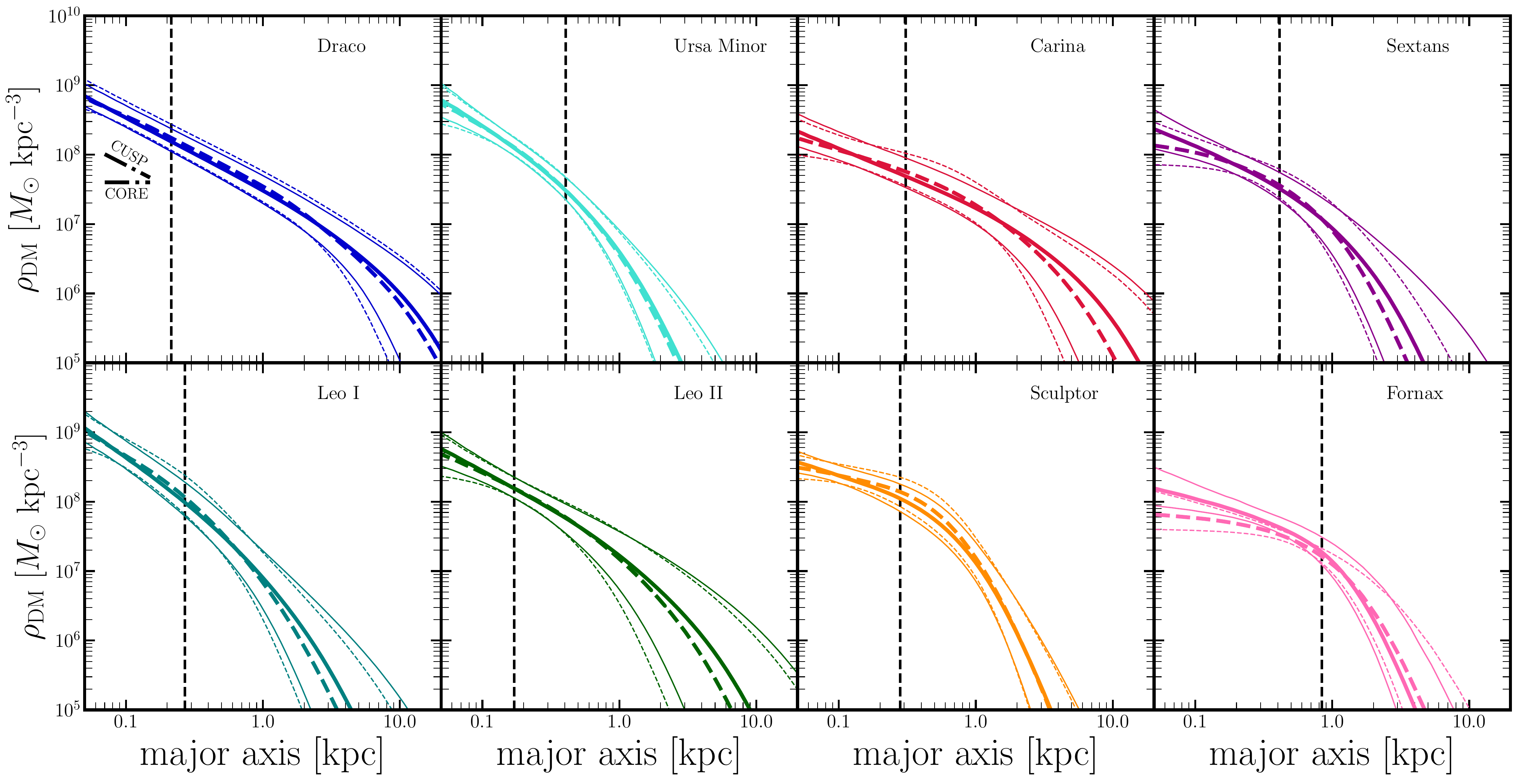}
    \caption{Dark matter density profiles of all dSphs, with taking into account a wider parameter range of $\gamma$~(described in Section~\ref{sec:bestdmprof}).
    The solid lines in each panel denote the median values~(thick) and the 68~per~cent confidence intervals~(thin) calculated by our default parameter range ($0\leq\gamma\leq2.5$), while the dashed ones are calculated by a new parameter range ($-2.5\leq\gamma^{\prime}\leq2.5$, but if $\gamma^{\prime}<0\rightarrow\gamma=0$).
    The vertical dashed lines in each panel correspond to their half-light radii.}
    \label{dmpro_bias}
\end{figure*}

In order to demonstrate the robustness of our results, especially regarding the inner slope of a dark matter density profile, $\gamma$,
we show the case when a wide range of prior for $\gamma$ is adopted, compared to our fiducial parameter range of $\gamma$ ($0\leq\gamma\leq2.5$).
Namely, we show here the case of a flat prior over range $-2.5\leq\gamma^{\prime}\leq2.5$, and we impose $\gamma=0$ if $\gamma^{\prime}$ has a negative value and $\gamma=\gamma^{\prime}$ otherwise.
This is because the fiducial parameter range of $\gamma$ ($0\leq\gamma\leq2.5$) might lead to a bias toward cuspy density profiles.
Using this new prior, we re-run the same MCMC fitting procedure described in Section~\ref{sec:fitting}.
Figure~\ref{dmpro_bias} shows the comparison of the inferred dark matter density profiles for all sample dSphs for the fiducial~(solid) and wider~(dashed) prior ranges.
The thick and thin lines in each panel denote the median and the 68~per~cent confidence intervals.
It is found from this figure that the galaxies having a cusped dark matter halo like Draco and Ursa Minor are not so much affected by new prior, whilst the effect of new prior makes Fornax and Sextans less dense core.
Therefore, we bear in mind that Fornax and Sextans are possible to have a cored dark matter density.
On the other hand, we can confirm that our results for Draco and Ursa~Minor have cusped dark matter halos.

\begin{figure*}[t!]
	\includegraphics[scale=0.4]{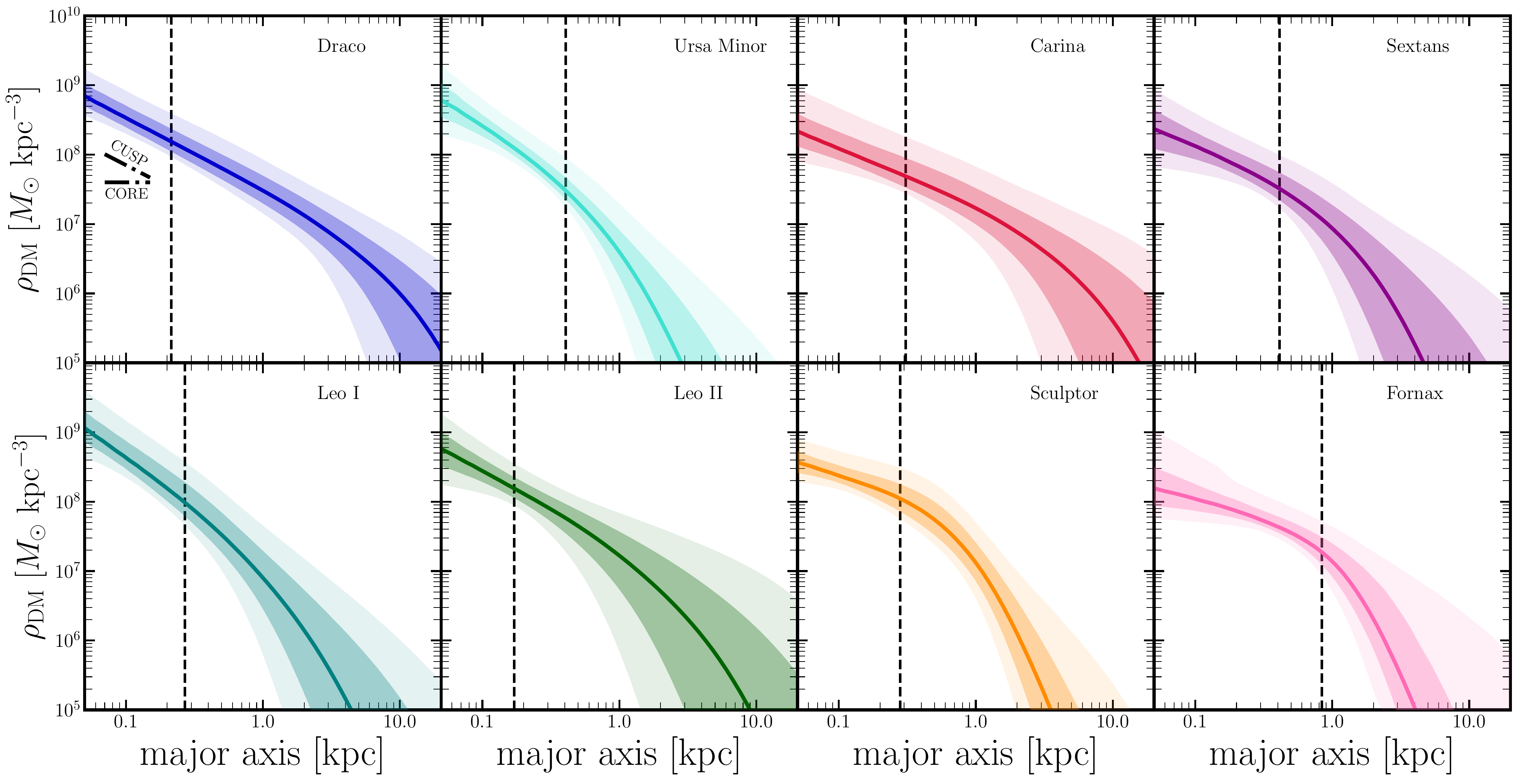}
    \caption{Dark matter density profiles along major axes of the galaxies derived from our Jeans analysis.
    The solid line in each panel denotes the median value, and the dark and light shaded regions denote the 68 and 95~per~cent confidence intervals.
    The vertical dashed line in each panel corresponds to the half-light radius of each galaxy.
    In the panel for Draco, we mark on two power law density profiles, $\rho_{\rm DM}\propto r^{-1}$ (cusp) and $\rho_{\rm DM}={\rm const.}$ (core) under the shaded regions.}
    \label{dmpro}
\end{figure*}

\subsection{Astrophysical factors} \label{sec:bestjd}
The Galactic dSphs are promising targets for indirect searches for particle dark matter through $\gamma$-rays or X-rays stemmed from annihilating and decaying dark matters~\citep[e.g.,][]{1978ApJ...223.1015G,2012AnP...524..479B}, because they contain a good deal of dark matter with low astrophysical backgrounds and are located at relative proximity.
The signal flux of the dark matter annihilation or decay depends only on two important factors.
One is the particle physics factor which is based on the microscopic physics of particle dark matter, while another is the astrophysical factor derived by line-of-sight integrals over the dark matter distribution within the system.
The latter largely depends on the estimate of the signal flux.
Therefore, an accurate estimation of the astrophysical factor in the dSphs is of crucial importance so that we can set robust constraints on the particle nature of dark matter candidates.

Previous works have estimated the astrophysical factors for these galaxies considering various uncertainties: the spatial dependence of stellar velocity anisotropy~\citep{2016JCAP...07..025U}, non-sphericity of a dark matter distribution~\citep{2015MNRAS.446.3002B,2016MNRAS.461.2914H,2017PhRvD..95l3012K}, halo truncation radius~\citep{2015ApJ...801...74G}, prior bias of Bayesian analysis~\citep{2009JCAP...06..014M}, and foreground contamination of stars~\citep{2016MNRAS.462..223B,2017MNRAS.468.2884I,2018MNRAS.479...64I,2020arXiv200204866H}.

Here we calculate the astrophysics factors of the dSphs focusing only on non-sphericity based on the generalized Hernquist density profile of their dark matter halos.
In fact, (sub-) subhalos and substructures can boost the annihilation signals~\citep[subhalo boost,][]{2017MNRAS.466.4974M,2018PhRvD..97l3002H,2020MNRAS.492.3662I}. However, this boost contributes little to the signals on the dSph's mass scales, and thus we do not include this boost to estimate $J$-factor values.
To compare with previous works, we show only the factors integrated within a fixed solid angle $0.5^{\circ}$.

The astrophysical factors are written as 
\begin{eqnarray}
J &=& \int_{\Delta\Omega}\int_{\rm los}d\ell d\Omega\rho^2_{\rm DM}(\ell,\Omega) \ \hspace{5mm}  [{\rm annihilation}], \\
D &=& \int_{\Delta\Omega}\int_{\rm los}d\ell d\Omega\rho_{\rm DM}(\ell,\Omega) \ \hspace{5mm}  [{\rm decay}],
\end{eqnarray}
which are so-called $J$- and $D$-factors, defined as the integrated dark matter density squared for annihilation and the dark matter density for decay, respectively, over a distance $\ell$ along a line-of-sight and a solid angle $\Delta\Omega$.
Using these equations, we estimate the median and its uncertainties of the astrophysical factors from the posterior PDFs of the dark matter halo parameters.

\begin{figure}
	\begin{center}
	\includegraphics[width=\columnwidth]{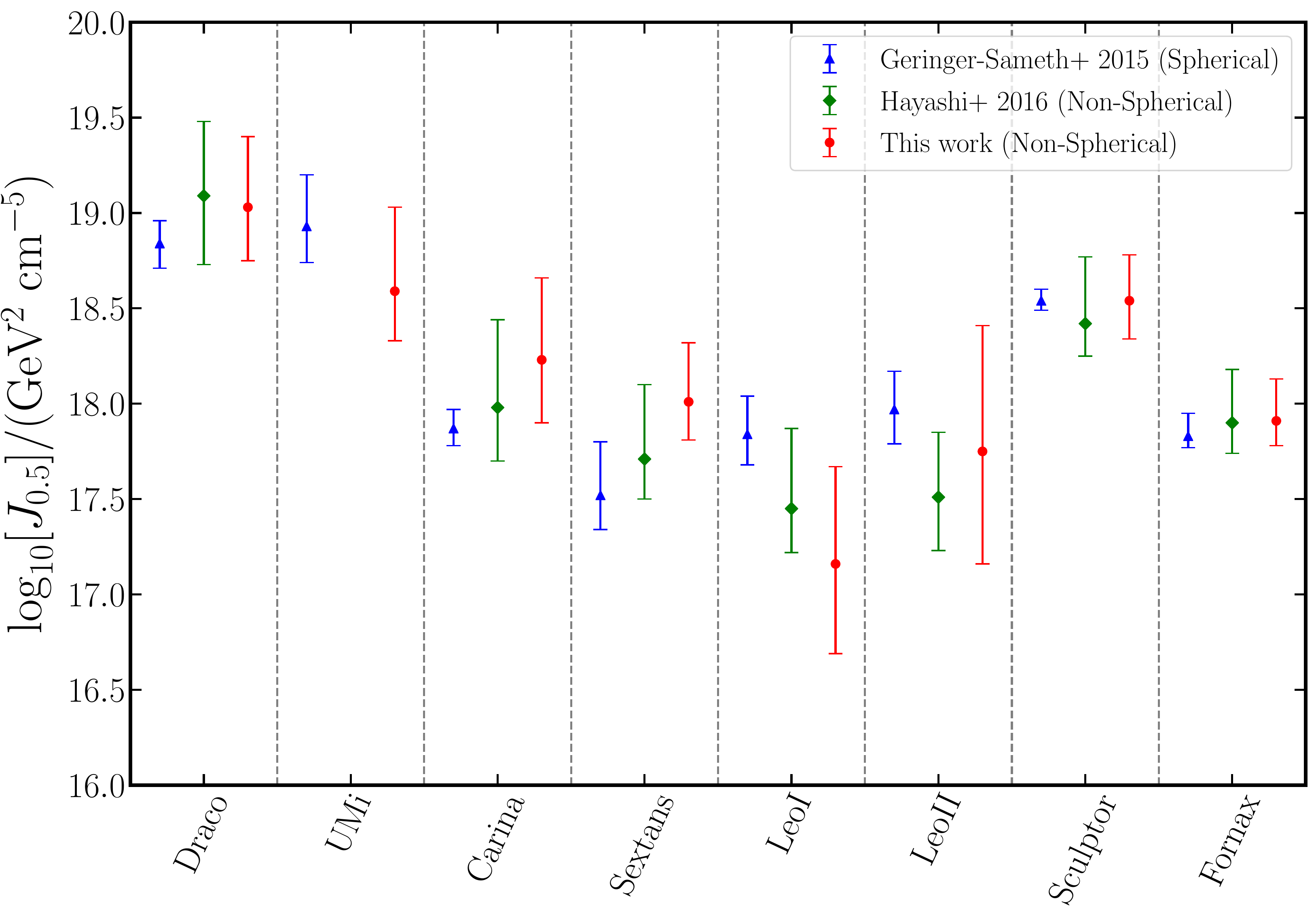}
	\includegraphics[width=\columnwidth]{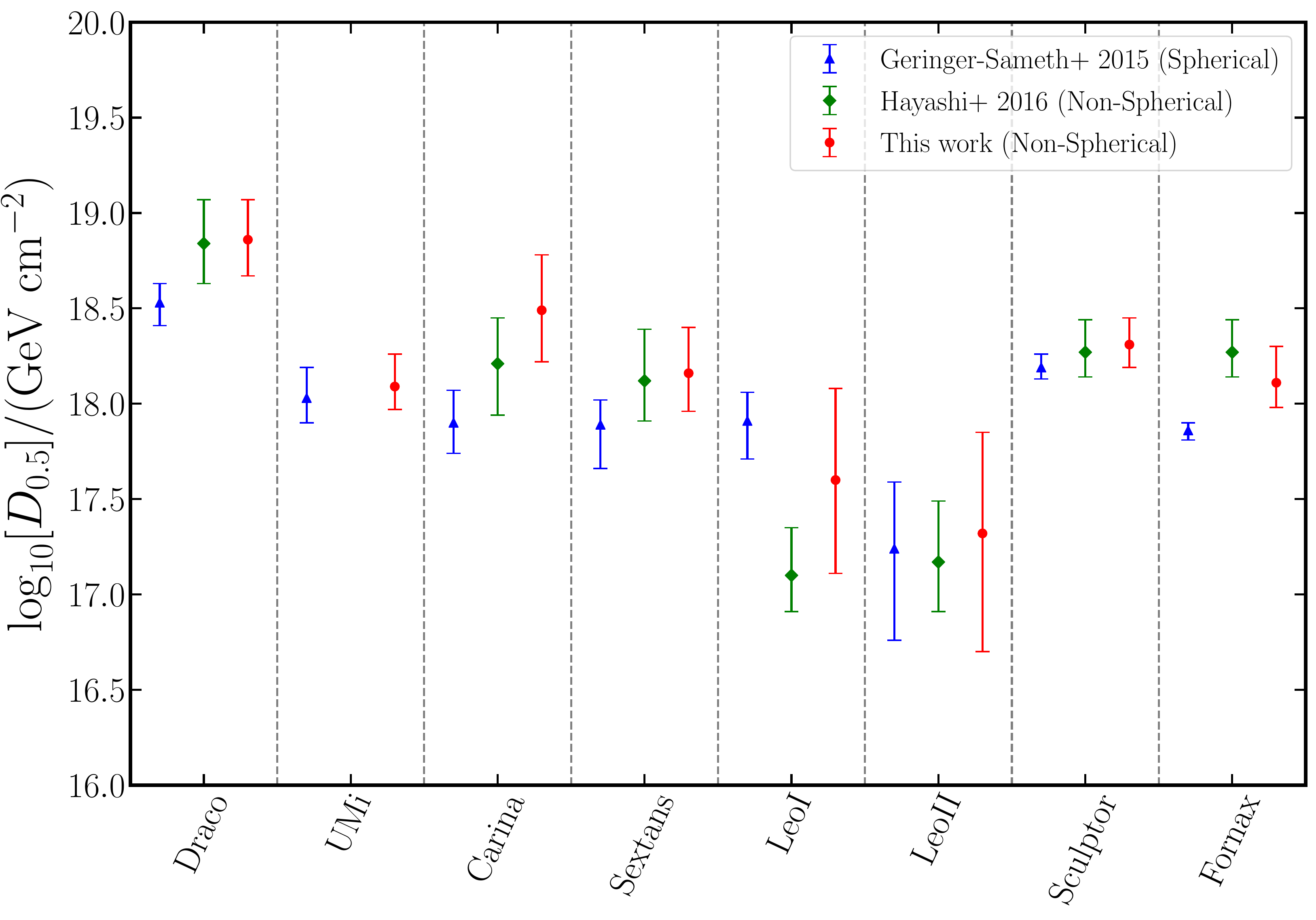}
	\end{center}
    \caption{Comparison of $J_{0.5}$~(top) and $D_{0.5}$~(bottom) calculated from previous and this works.
    The blue and green symbols are estimated by~\citet{2015ApJ...801...74G} and \citet{2016MNRAS.461.2914H}.
    The red symbols denote the results of this work.}
    \label{JDcomp}
\end{figure}

Table~\ref{table3} shows the $J$ and $D$ values integrated within $\Delta\Omega=0.5^{\circ}$ of our results.
Figure~\ref{JDcomp} displays a comparison of the $J$ (top) and $D$ (bottom) values of our results with those of previous works.
In this figure, the red colored points with error bars are the median values in this work with 68~per~cent confidence intervals.
The blue ones denote these values reported by \citet{2015ApJ...801...74G}, which assumed a spherical dark matter halo with a generalized Hernquist density profile and performed Jeans analysis.
The green ones are evaluated by \citet{2016MNRAS.461.2914H}, which assumed an axisymmetric dark matter halo.
The differences between them in this figure are caused primarily by the assumption of shapes of dark matter halos (spherical or non-spherical) as already discussed by \citet{2016MNRAS.461.2914H} and dark matter density profiles.
The latter means that \citet{2016MNRAS.461.2914H} imposed that the outer slope of dark matter profiles is $\rho\propto r^{-3}$ and the sharpness parameter $\alpha$ in Equation~\ref{DMH} is fixed at $2$ for simplicity, while the dark matter profiles in this work and \citet{2015ApJ...801...74G} take into account these parameter as free parameters.

\input{table3}

From Figure~\ref{JDcomp}, we conclude that because of having a cuspy dense dark matter halo and of the close distance to the Sun, Draco is the most promising detectable target for an indirect search of dark matter annihilation and decay among all sample dSphs.

\section{Discussion} \label{sec:discussion}
\subsection{Comparison dark matter profiles with previous works}
In this section, we compare our estimated dark matter density profiles with other works based on different methods or assumptions.

\citet{2019MNRAS.484.1401R} considered non-parametric dynamical mass models based on a spherical Jeans equation, {\sc GravSphere}~\citep{2017MNRAS.471.4541R,2018MNRAS.481..860R} to measure the dark matter density profiles of dwarf spheroidal/irregular galaxies, and then they found the relation between the central densities of dark matter halos and the stellar vestiges of galaxy evolution such as a star formation history, stellar mass, and stellar-to-halo mass ratio.
Regarding the inner slopes of dark matter density profiles in the dSphs, they showed that Draco favors a cusped dark matter halo which is consistent with an NFW profile, while Fornax has a shallower inner density profile $\gamma\sim0.3$.
This trend is similar to that in this work.
They mentioned, however, the caveat that the estimation of an inner slope of a dark matter profile using their method is largely affected by a choice of priors.
Therefore, they utilized a dark matter density within 150~pc, $\rho_{\rm DM}(150\ {\rm pc})$, to discuss a diversity of the central dark matter densities in the dwarf galaxies, instead of their inner slopes. 
We also discuss $\rho_{\rm DM}(150\ {\rm pc})$ calculated by our models and then find that this physical quantity is useful to understand the dynamical evolution of dark matter halos in the Universe.
We discuss them further in the following subsection.

Owing to recent spectroscopic observations for the dSphs, some of them have multiple stellar populations, in which the metal-rich stars are centrally concentrated and have colder kinematics, while the metal-poor ones are more extended and have hotter kinematics~\citep[e.g.,][]{2006A&A...459..423B,2008ApJ...681L..13B}.
Using the coexistence of such multiple populations, \citet{2011ApJ...742...20W} statistically separated multiple stellar components by applying their constructed likelihood function for spatial, metallicity, and velocity distributions of the stars, and then inferred the slopes of dark matter densities of Sculptor and Fornax.
They concluded that both galaxies have cored dark matter halos and a cuspy profile can be ruled out with high statistical significance.
However, this method imposes that both stellar and dark matter distributions are spherical symmetric.
This sphericity can accompany a systematic bias, and an inner slope inferred by this method depends largely on viewing angles~\citep{2013MNRAS.431.2796K,2013MNRAS.433L..54L,2018MNRAS.474.1398G}.

\citet{2012ApJ...754L..39A} and \citet{2013MNRAS.429L..89A} applied the projected virial theorem to these multiple stellar components for Sculptor and Fornax, respectively and concluded that these dSphs do not have cusped dark matter profiles.
On the other hand, using these multiple populations, several other works concluded that Sculptor has a cusped dark matter halo based on a phase space distribution function method~\citep{2017ApJ...838..123S}, whilst it is difficult to distinguish between cusp and core based on a Schwarzschild method~\citep{2013MNRAS.433.3173B} and Multi-Gaussian expansion model~\citep{2016MNRAS.463.1117Z}.
Although the dark matter inner slopes in Fornax and Sculptor are still under debated, those inferred by our mass models prefer to be less cuspy than an NFW profile.

Axisymmetric dynamical models based on Schwarzschild technique have been developed and applied to the kinematic data of the dSphs~\citep{2012ApJ...746...89J,2013ApJ...775L..30J,2013ApJ...763...91J}.
\citet{2013ApJ...763...91J} applied these models to the data of Draco and found that its dark matter inner slope is consistent with an NFW profile.
This agrees well with our mass models for Draco.
\citet{2013ApJ...775L..30J} performed the same analysis with respect to the other classical dSphs (Carina, Fornax, Sculptor and Sextans) and concluded that these galaxies have an unified cusped profile, but there are considerable large uncertainties.

\begin{figure*}[t]
	\begin{center}
	\includegraphics[scale=0.9]{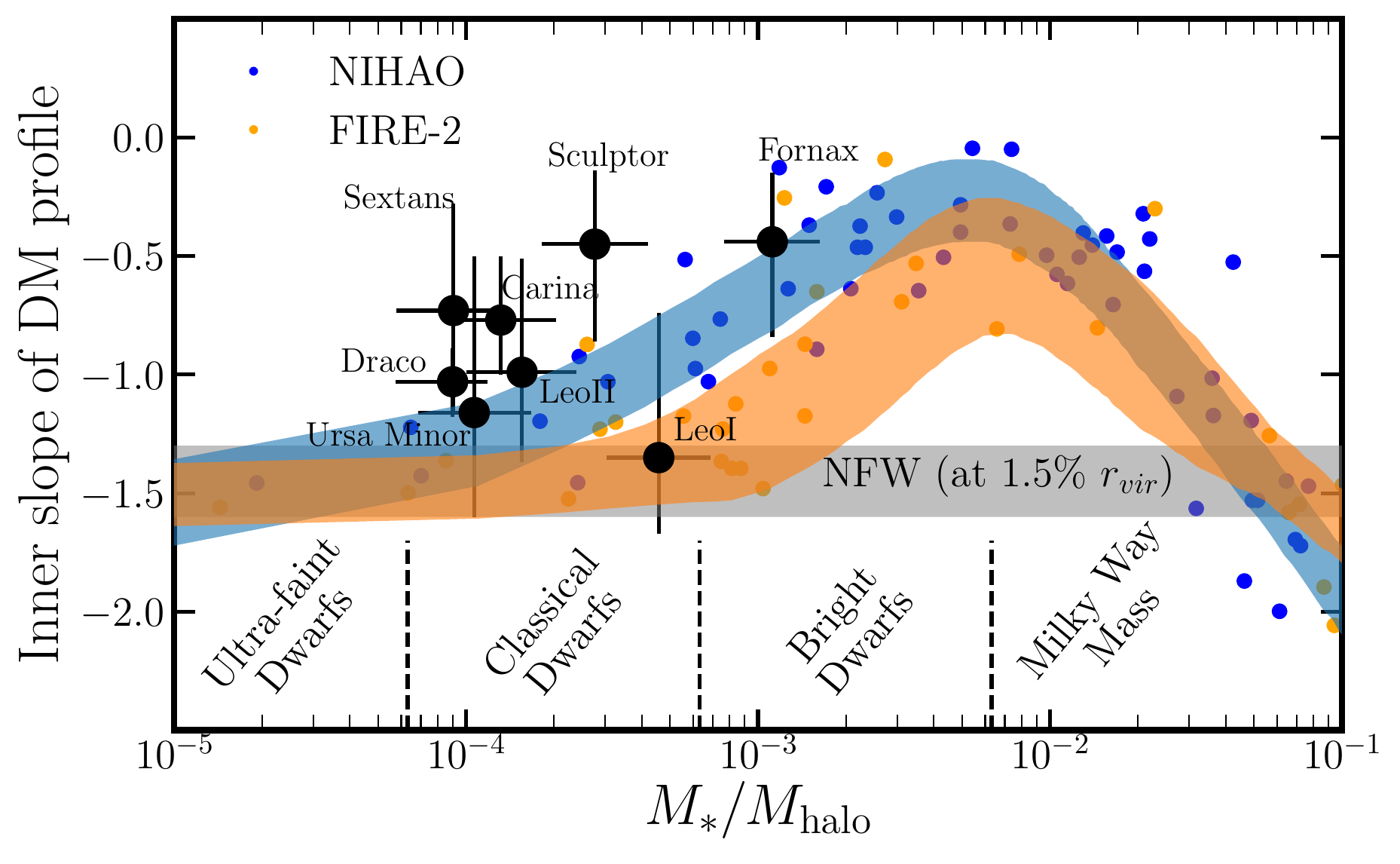}
	\end{center}
    \caption{The impact of baryonic feedback on the inner profiles of dark matter halos. 
    The inner dark matter density slope at $1.5$\%$R_{\rm vir}$ is shown as a function of the ratio of stellar-to-halo masses.
    The filled black circles with error bars are the results from this work. 
    The shaded gray band shows the expected range of dark matter profile slopes for NFW as derived from dark matter only simulations~\citep{2016MNRAS.456.3542T}.
    The blue and orange points are expected from NIHAO~\citep{2016MNRAS.456.3542T} and FIRE-2~\citep{2017MNRAS.471.3547F,2018MNRAS.480..800H} hydrodynamical plus dark mater simulations, respectively.
    The blue and orange shaded bands are the expected range from NIHAO~\citep{2016MNRAS.456.3542T} and FIRE-2~\citep{2020arXiv200410817L} predictions, respectively (to guide the eye).}
    \label{GammaMsMh}
\end{figure*}

\subsection{The origin of a diversity of inner dark matter slopes} \label{sec:diversity}
In Figure~\ref{dmpro}, we show that the classical dSphs have a wide range of central dark matter density profiles.
In this section, we discuss what the origin of this diversity is.
To this end, we investigate the relation between the central dark matter density profiles and stellar properties of the dSphs.

\subsubsection{Inner dark matter density slope versus stellar-to-halo mass ratio}
Recent dark matter plus hydrodynamical simulations have shown that an inner slope of a dark matter density profile depends largely on the ratio of stellar mass to total halo mass.
Figure~\ref{GammaMsMh} shows the logarithmic slope of the dark matter density profile at $1.5$\% of the virial radius, $R_{\rm vir}$, as a function of the ratio of stellar-to-halo masses, $M_{\ast}/M_{\rm halo}$, predicted from NIHAO~\citep[][magenta]{2016MNRAS.456.3542T} and FIRE-2~\citep[][cyan]{2017MNRAS.471.3547F,2018MNRAS.480..800H} simulations.
Note that baryon feedback for bright dwarf galaxies ($\log_{10}(M_{\ast}/M_{\rm halo})\sim-3$~to~$-2$) has a systematic impact on inner slopes, while for the fainter galaxies with $\log_{10}(M_{\ast}/M_{\rm halo})\lesssim-3.5$, the impact of baryonic feedback is negligible. 
Therefore, these simulations predict that the efficiency of baryonic feedback for a dark matter halo can provoke the diversity of dark matter inner slopes.

To test this prediction, we derive the relation between the dark matter inner slopes and $M_{\ast}/M_{\rm halo}$ for the current sample of dSphs, which is shown in Figure~\ref{GammaMsMh}.
In order to calculate the ratio of stellar-to-halo masses, we employ the self-consistent abundance matching model computed by \citet{2013MNRAS.428.3121M} and adopt the stellar masses of the dSphs taken from \citet{2012AJ....144....4M}.
The filled black circles with error bars in Figure~\ref{GammaMsMh} show the results of the classical dSphs inferred by our analysis.
Although there are still large uncertainties in both the inner slopes and the stellar-to-halo mass ratios, 
the systematic trend in the plots is generally in agreement with the predictions from recent numerical simulations, which are presented in blue~(NIHAO: \citealt{2016MNRAS.456.3542T}) and orange (FIRE-2: \citealt{2020arXiv200410817L}) shaded region in the figure.
To make an attempt to characterize the trend quantitatively, we employ a least squares fitting method to
determine the slope of $\gamma$ as a function of $M_{\ast}/M_{\rm halo}$, and we find $\gamma\propto\log_{10}(M_{\ast}/M_{\rm halo})^{0.27\pm0.15}$.
Thus, we confirm that $\gamma$ is slightly proportional to $M_{\ast}/M_{\rm halo}$ on dwarf-galaxy scales.

However, comparing between these shaded bands in detail, there is a systematic difference especially on classical dwarf galaxy scales stemmed from the different prescriptions of hydrodynamics regime.
Thus, the predicted relation between dark matter inner slope and stellar-to-halo mass ratio still has large uncertainties.
Regarding this relation, \citet{2019arXiv191100544K} have argued that a self-interacting dark matter (SIDM) model combined with the impact of a baryon potential on the halo profile can also reproduce the diversity of the inner dark matter density profiles for low surface brightness galaxies.
However, the corresponding $M_{\ast}/M_{\rm halo}$ in these galaxies are greater than $-3$, and it is thus unclear for the diversity in the current fainter dwarf galaxy scales.  

We also investigate the relation between the inner density slopes and their stellar masses and the orbital properties of the dSphs~(apocenter radius, orbital eccentricity, angular momentum, the time elapsed since the last apocenter and pericenter) but we find no clear relations.

\begin{figure}[t!]
	\begin{center}
	\includegraphics[scale=0.3]{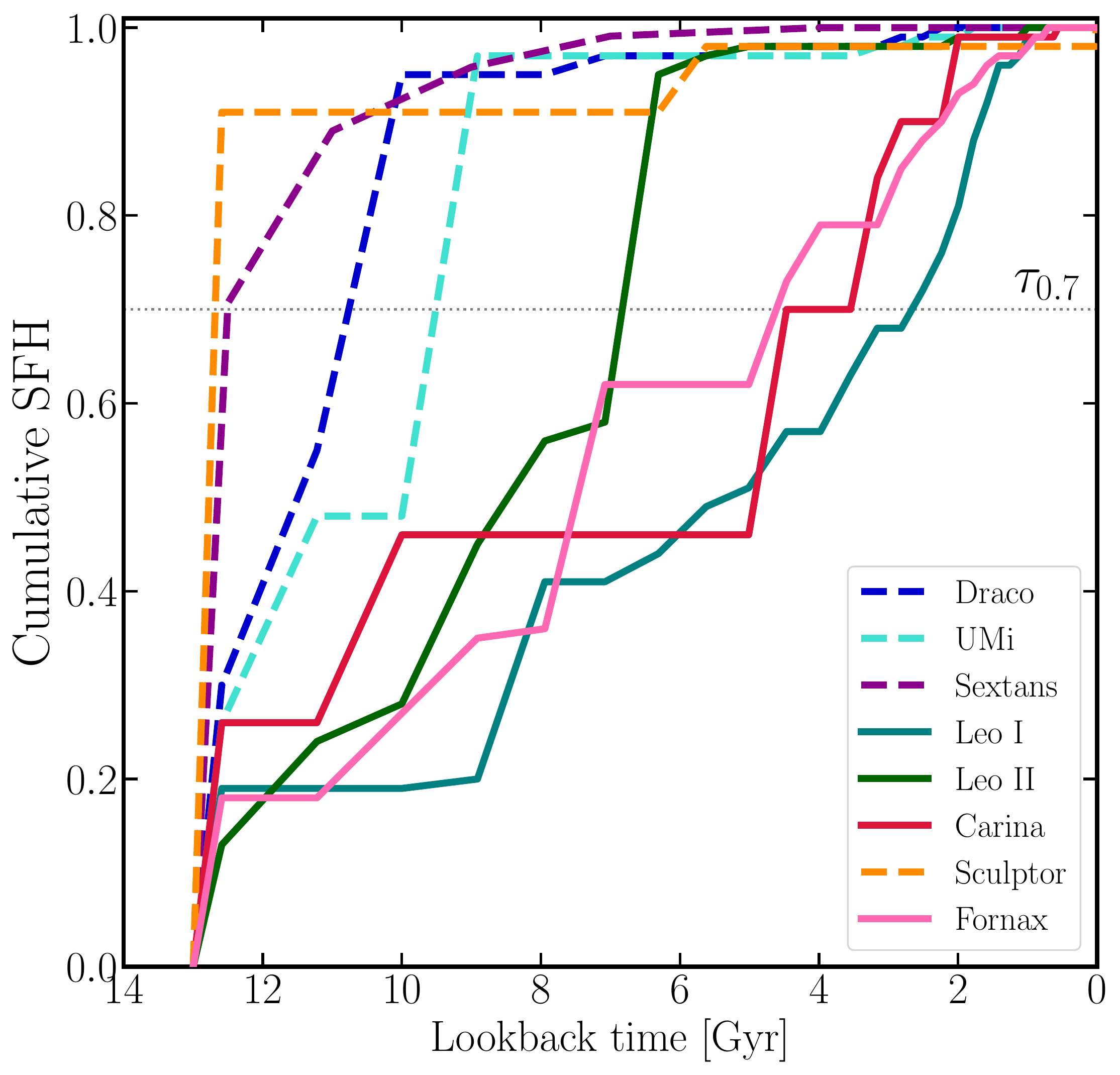}
	\includegraphics[scale=0.3]{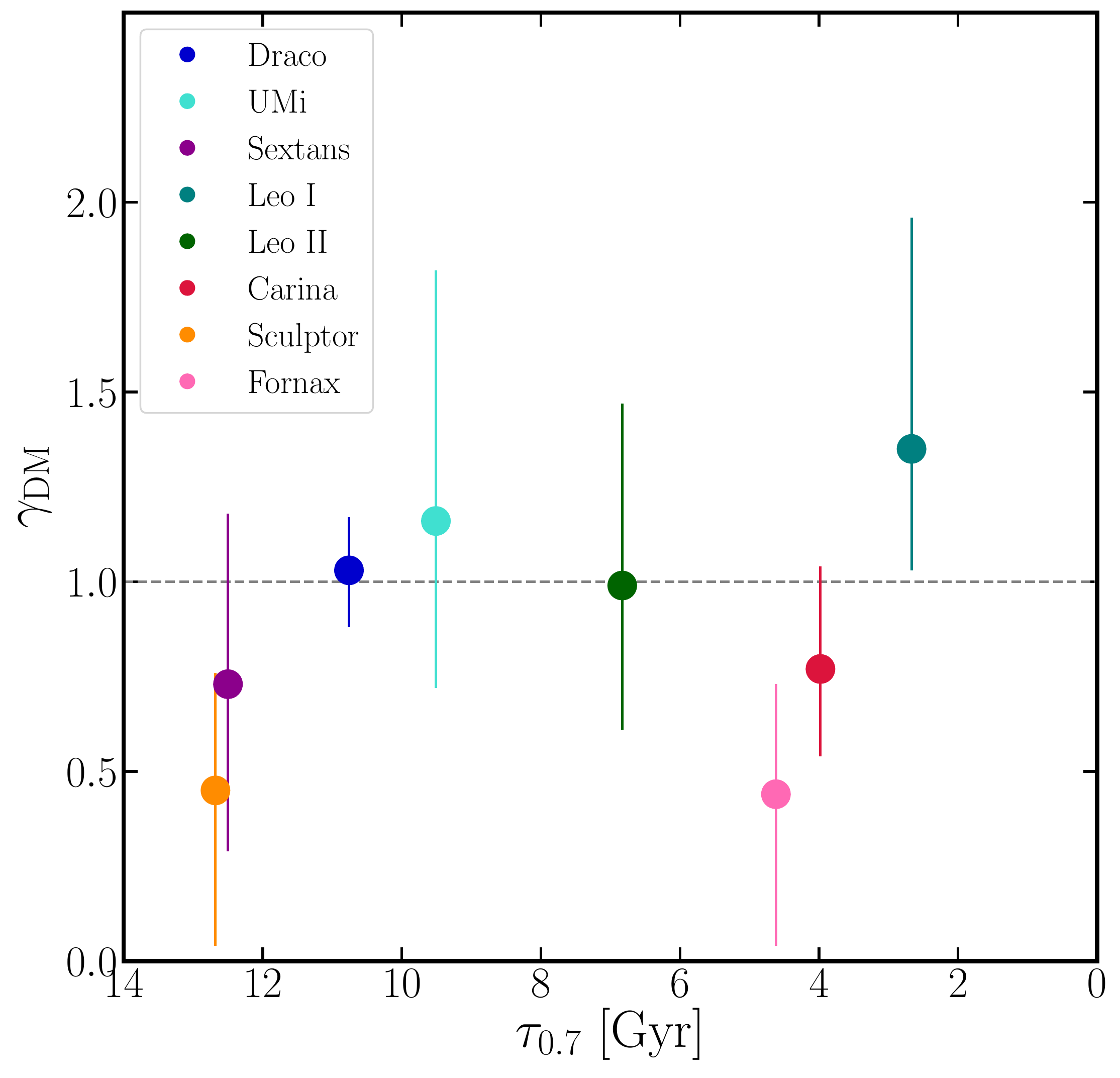}
	\end{center}
    \caption{{\it Top panel}: Cumulative star formation history of dwarf satellites in the MW taken from \citet{2009ApJ...703..692L} for Sextans and \citet{2014ApJ...789..147W} for the other classical dSphs.
    {\it Bottom panel}: Inner slope parameter of dark matter density profile $\gamma$ as a function of the lookback time of achieving 70~per~cent of current stellar masses, $\tau_{0.7}$.}
    \label{Gamma_esn}
\end{figure}

\subsubsection{Inner dark matter density slope versus SFH}
The relation in Figure~\ref{GammaMsMh} implies that an inner dark matter slope depends on stellar feedback associated with star formation activity.
Indeed, some high-resolution dark matter and hydrodynamical simulations have shown an inner slope of a dark matter density profile depends on star formation history~(SFH)~\citep[e.g.,][]{2014ApJ...789L..17M, 2015MNRAS.454.2092O}.
In particular, \citet{2015MNRAS.454.2092O} predicted that the dwarfs with rapid SFHs tend to have cuspy dark matter density profiles, while ones with consecutive SFHs have cored ones at the present day.
Therefore, we investigate whether this dependence indeed exists by comparing it with the observed SFH of dSphs.

To this end, we adopt the SFHs derived by \citet{2009ApJ...703..692L} for Sextans and \citet{2014ApJ...789..147W} for the other classical dSphs.
The top panel in Figure~\ref{Gamma_esn} displays the cumulative SFHs of the classical dSphs taken from their works.
As is shown in the panel, the SFHs of the dSphs can be classified into two groups:
the dwarfs~(the dashed lines in the panel) that formed the majority of their stellar component early on (before $z\simeq2$), and the other ones~(the solid ones) that formed only a small fraction of their stars at early times and continued forming stars over almost a Hubble time~\citep{2015ApJ...811L..18G,2018MNRAS.479.1514B}.
To quantify these properties of the dwarfs, we estimate the lookback time at achieving 70~per~cent of the current stellar mass of these dSphs, $\tau_{0.7}$ (as indicated as a black horizontal dotted line in the left panel in Figure~\ref{Gamma_esn}).
$\tau_{0.7}$ can characterize the duration and efficiency of star formation in dSphs.
The bottom panel in Figure~\ref{Gamma_esn} shows the comparison between $\tau_{0.7}$ and dark matter inner slope, $\gamma$, from our analysis.
According to the prediction from \citet{2015MNRAS.454.2092O}, we expect that the galaxies with higher $\tau_{0.7}$ may have cuspy dark matter density profiles.
From this figure, however, we find no clear relation between them within uncertainties of $\gamma$.
Therefore, the diversity of the dark matter inner slopes cannot be explained straightforwardly by SFH within the current observation and model uncertainties.
One of the possible reasons why there is no relation could be that the cusp-core transition requires the resonance between dark matter particles and a gas density oscillation induced by periodic SN feedbacks.
\citet{2014ApJ...793...46O} suggested that to transform cusp into core, at least 50 oscillations with $\mathcal{O}(100)$~Myr periods are needed.
Unfortunately, current photometric and spectroscopic observations are difficult to resolve such a oscillatory star formation activity.

\subsubsection{Dark matter density at 150~pc}
\citet{2019MNRAS.484.1401R} proposed to use the dark matter density at a common radius of 150~pc from the center of each galaxy, $\rho_{DM}(150\ {\rm pc})$, which is insensitive to the choice of a $\gamma$'s prior in spherical mass models.
Using this density, \citet{2019MNRAS.490..231K} pointed out the anti-correlation between $\rho_{DM}(150\ {\rm pc})$ and their orbital pericenter distances, $r_{\rm peri}$, of the classical dSphs.
This implies a survivor bias which means that galaxies with low dark matter densities were completely destroyed by strong tidal effects.
Following these works, we also calculate the dark matter density at 150~pc along the major axis of the sample dSphs, considering the non-sphericity of a dark matter halo, and the calculated $\rho_{DM}(150\ {\rm pc})$ are tabulated in the last column of Table~\ref{table2}. 

First, we compare their $\rho_{DM}(150\ {\rm pc})$ to stellar masses and stellar-to-halo mass ratios.
\citet{2019MNRAS.484.1401R} presented the anti-correlation between them, but we do not find clear relations of $\rho_{DM}(150\ {\rm pc})$-$M_{\ast}$ and $\rho_{DM}(150\ {\rm pc})$-$M_{\ast}/M_{\rm halo}$.
This is caused by the fact that \citet{2019MNRAS.484.1401R} discussed these correlations by including not only the dSphs but dwarf irregular galaxies which have HI gas rotation curves.
These gas-rich galaxies have higher $M_{\ast}$ and $M_{\ast}/M_{\rm halo}$ and much lower $\rho_{DM}(150\ {\rm pc})$ than those of the dSphs, thereby the galaxies make the correlations conspicuous.

\begin{figure}[t!]
	\begin{center}
	\includegraphics[width=\columnwidth]{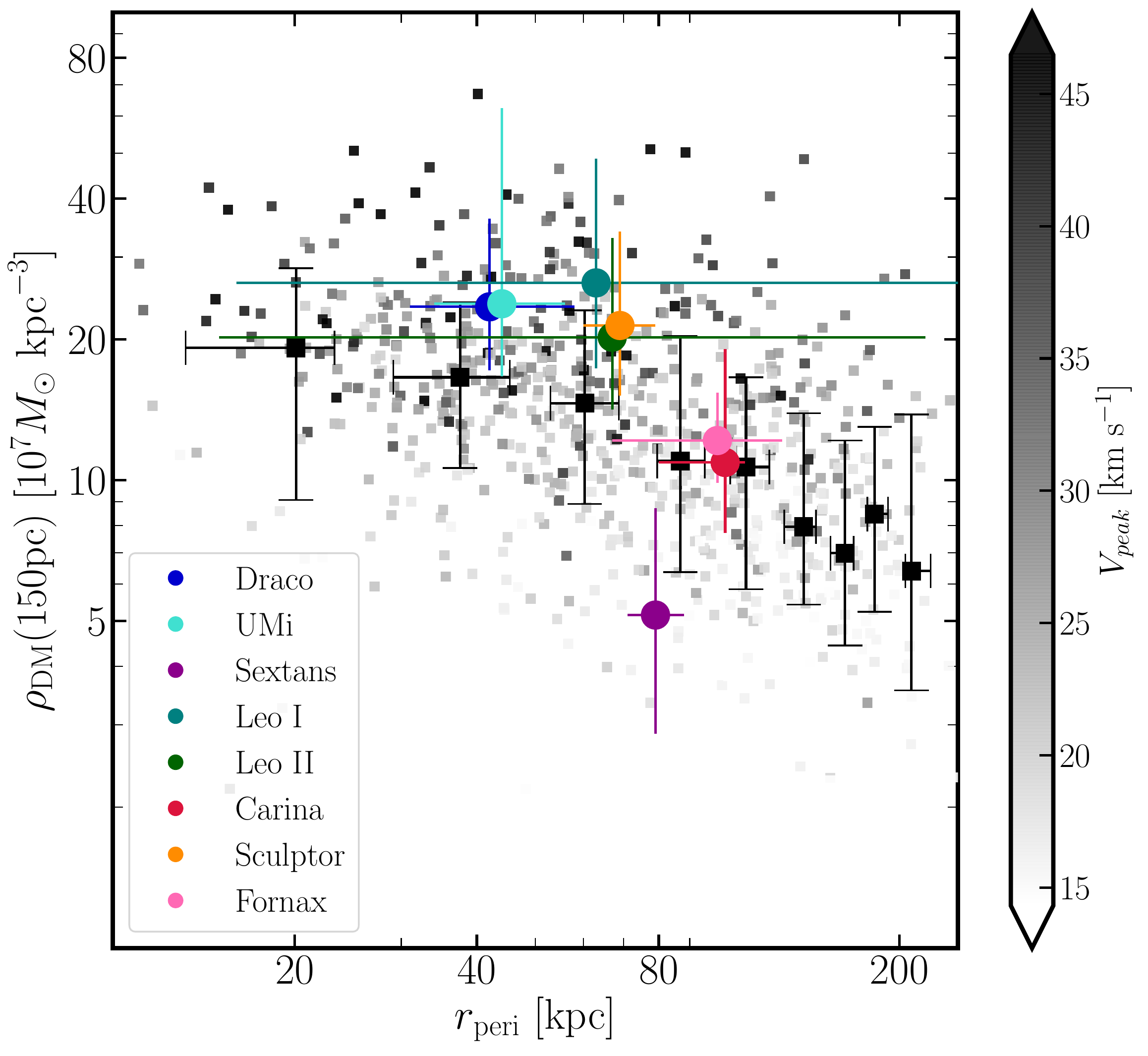}
	\end{center}
    \caption{Dark matter densities at 150~pc, $\rho_{DM}(150\ {\rm pc})$, versus pericenter radii, $r_{\rm peri}$, of the dSphs.
    The colored filled circles with error bars are the classical dSphs from our Jeans analysis.
    The filled small squares are the individual subhalos predicted from dark matter simulations~(Ishiyama et al. in prep.).
    The gray scale indicates the maximum circular velocities of subhalos over their formation histories~(the redshift when they were first accreted on to a host).
    The big black filled squares with error bars are the stacked $\rho_{DM}(150\ {\rm pc})$ and $r_{\rm peri}$ in each radial bin.
    The error bars correspond to the 16th and 84th percentiles of the subhalos in each bin.}
    \label{rho150_rperi}
\end{figure}

\begin{figure*}[t!]
	\includegraphics[scale=0.43]{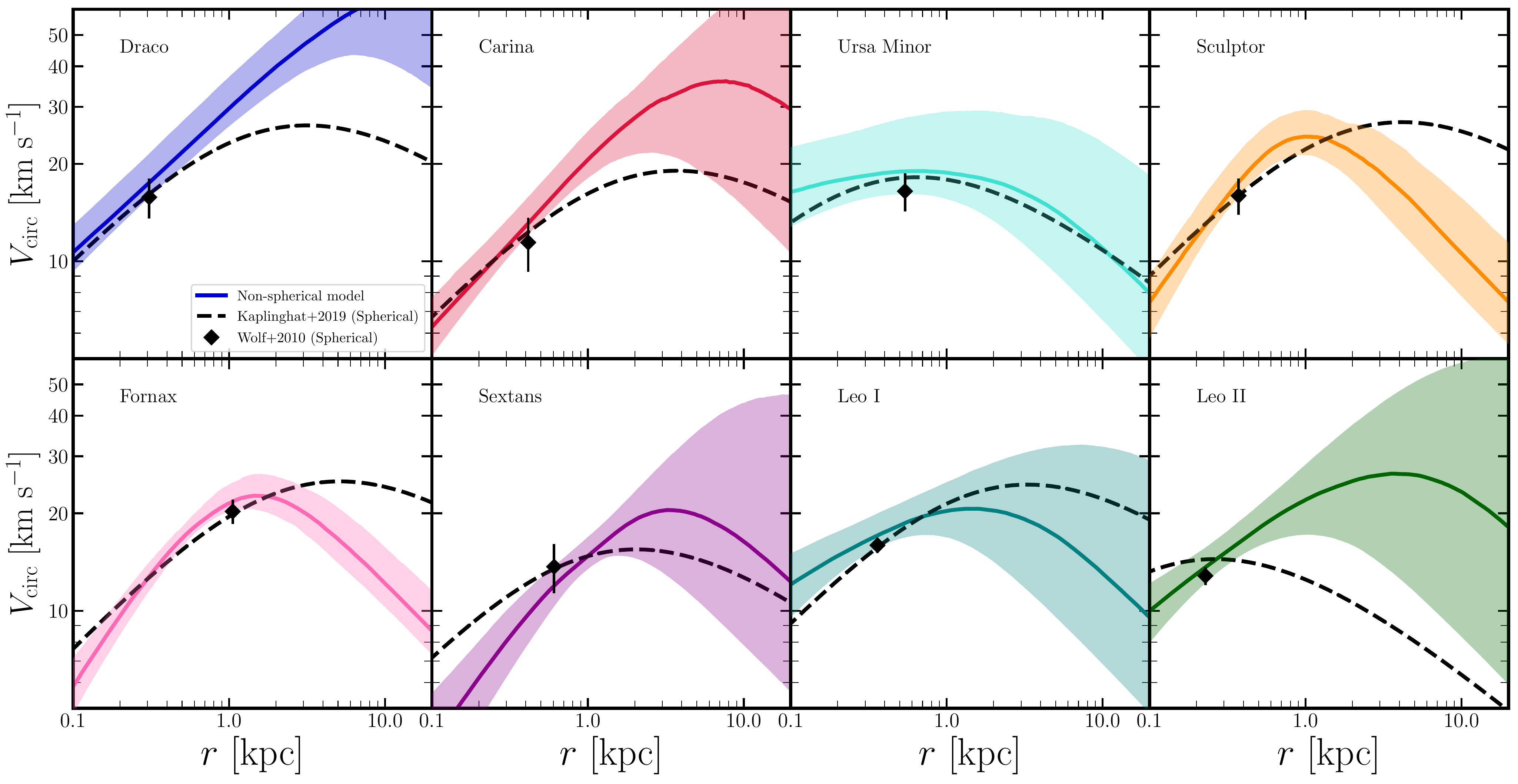}
    \caption{The circular velocity profiles for all sample dSphs.
    The colored solid line and shaded band in each panel show median and the 68~percent confidence intervals calculated by our non-spherical mass models.
    The dashed lines depict the results from spherical mass models assuming NFW cusped dark matter density profiles taken from~\citet{2019MNRAS.490..231K}.
    The black diamonds correspond to the mass estimator of~\citet{2010MNRAS.406.1220W}.}
    \label{vcirc_all}
\end{figure*}

Second, we investigate the anti-correlation between $\rho_{DM}(150\ {\rm pc})$ and $r_{\rm peri}$.
For the pericenter radius, we adopt the values presented by \citet{2018A&A...619A.103F}, which estimated using the recent Gaia data~\citep{2018A&A...616A..12G} and assuming a Milky Way potential model with mass of $0.8\times10^{12}M_{\odot}$.
Figure~\ref{rho150_rperi} shows the relation between $\rho_{DM}(150\ {\rm pc})$ and $r_{\rm peri}$.
The colored filled circles with error bars are the inferred $\rho_{DM}(150\ {\rm pc})$ of the sample dSphs.
From this plot, we find the anti-correlation between them similar to \citet{2019MNRAS.490..231K}, even though there are still large uncertainties.

We also compare with dark matter subhalos predicted from dark matter only simulations.
In this work, we utilize a high resolution $N$-body simulation, named Phi-4096,  performed by~\citet{2020arXiv200714720I}.
The detail of the simulation is as below.
Using a massively parallel TreePM code
GreeM~\footnote{http://hpc.imit.chiba-u.jp/~ishiymtm/greem/}
~\citep{2009PASJ...61.1319I,2012arXiv1211.4406I},
we simulated the motion of $4096^3$ dark matter particles
in a comoving box with the side length of 16$ \, h^{-1} \rm Mpc$,
which corresponds to $5.13 \times 10^{3} \, h^{-1} M_{\odot}$ particle mass.
The gravitational softening length is 60 comoving $ \, h^{-1} \rm pc$.
The initial condition was constructed using the MUSIC code \citep{2011MNRAS.415.2101H}.
The cosmological parameters of the simulation
are $\Omega_0=0.31$, $\lambda_0=0.69$, $h=0.68$, $n_s=0.96$, and
$\sigma_8=0.83$, which are consistent with the measurement of
cosmic microwave background by the Planck satellite~\citep{2018arXiv180706209P}.
To identify halos and subhalos and construct merger trees, we used
ROCKSTAR phase space halo/subhalo finder ~\citep{2013ApJ...762..109B}
and consistent trees code ~\citep{2013ApJ...763...18B}.  We picked up
Milky Way-sized host halos with the mass of 
$3.4 \times 10^{11} < M_{\rm vir} < 2.0 \times 10^{12} \, h^{-1} M_{\odot}$ at z=0, where $M_{\rm vir}$ is the
halo virial mass.
The total number of host halos is 27.

To compute $\rho_{DM}(150\ {\rm pc})$ of the simulated dark subhalos, we use the scale density and radius of each subhalo, supposing spherical NFW dark matter halos.
In Figure~\ref{rho150_rperi}, the small filled squares denote the predicted subhalos associated with these Milky Way-sized dark matter host halos, while the big black squares with error bars are the results from stacked analysis of the subhalos in each $r_{\rm peri}$ bin.
It is found from this plot that dark matter simulations indicate somewhat anti-correlation, and 
this correlation is similar to the observed one.
Moreover, we also find that the maximum circular velocities of subhalos over their formation histories, $V_{\rm peak}$, of subhalos depends slightly on $\rho_{DM}(150\ {\rm pc})$ and $r_{\rm peri}$.
In other words, the subhalos with higher $\rho_{DM}(150\ {\rm pc})$ and smaller $r_{\rm peri}$ (the left-top area in Figure~\ref{rho150_rperi}) tend to have large $V_{\rm peak}$.
Since subhalos with large $V_{\rm peak}$ were formed at earlier, most of them have dense central densities, and such subhalos can still survive even suffering from strong tidal effects.
In addition, we find that $\rho_{DM}(150\ {\rm pc})$ of subhalos depends on their host dark matter halo masses, which means that the subhalos associated with larger hosts have higher $\rho_{DM}(150\ {\rm pc})$ than those with smaller ones.
Therefore, this anti-correlation can be also dependence on a host halo mass.
Even though this anti-correlation seems to support a survivor bias suggested by \citet{2019MNRAS.490..231K}, we note that adding a stellar disk preferentially reduces the subhalo densities with smaller pericenter distances~\citep{2019MNRAS.490.2117R}, thereby we cannot make a final conclusion about the anti-correlation without considering the impact of the disk.

\subsection{Circular velocity profile}

$\Lambda$CDM theory has another serious problem that central densities of dark matter halos associated in the bright dSphs in Milky Way are significantly lower than those of the most massive subhalos in MW-sized halos in the $\Lambda$CDM simulations.
This problem is so-called the ``too-big-to-fail (TBTF)'' problem~\citep{2011MNRAS.415L..40B}.
In order to compare with the central densities in the observed and simulated dark matter halos, they adopted the maximum circular velocity, $V_{\rm max}$, for most massive ten subhalos.
On the other hand, for the observed ones, they used the circular velocities at the half-light radii of the dSphs, because this physical value is well-constrained by kinematic data~\citep{2010MNRAS.406.1220W}.
Instead of relying on such a single value of a circular velocity at a specific radius, we calculate a circular velocity profile directly from the posterior PDFs of the dark matter halo parameters.

In axisymmetric models, the circular velocity along a major axis can be calculated by
\begin{equation}
    V^2_{\rm circ}(R) = R\left| -\frac{\partial\Phi}{\partial R}\right|,
\end{equation}
where $\Phi$ is a gravitational potential originated from the dark matter density profile~ \citep{2008gady.book.....B}. 
The colored solid lines and shaded regions in Figure~\ref{vcirc_all} show the inferred circular velocity profiles for the classical dSphs from our models.
For comparison with our results, we also plot those profiles estimated by \citet{2019MNRAS.490..231K} and the circular velocities at their half-light radii of the dSphs, $V_{\rm circ}(r_{\rm half})$,~\citep{2010MNRAS.406.1220W}.
Interestingly, the both circular velocity profiles computed by axisymmetric and spherical models are consistent in the value of $V_{\rm circ}(r_{\rm half})$, but the shapes of these profiles, especially quantified with the values of $V_{\rm max}$, look quite different in different mass models.
This implies that $V_{\rm circ}(r_{\rm half})$ would not be an adequate tracer for comparison with the central densities in dark matter halos. 
However, there are huge uncertainties in our estimated circular velocities, especially their outskirts due to the lack of data sample.
Thus, a sufficient number of stellar kinematic sample out to their outer parts of the dSphs should be needed.

\section{Conclusion} \label{sec:conclusion}
In this paper, we revisit the core-cusp problem in the Galactic dSphs based on non-spherical Jeans analysis.
An advantage in these non-spherical models is that $\rho_{\rm DM}-\beta_{\rm ani}$ degeneracy occurred under the assumption of spherical symmetry can be mitigated.

Applying our non-spherical mass models to the latest kinematic data of the eight classical dSphs, we estimate their dark matter density profiles by marginalizing posterior distributions of dark matter halo parameters.
We find that most of these dSphs favor cusped or mildly cusped dark matter profiles in their centers rather than cored one.
In particular, Draco robustly has a cusped dark matter halo even considering a wide prior range.
Therefore, we conclude that there is no core-cusp problem in the classical dSphs.

We also find the diversity in the central dark matter density profiles.
Interestingly, this diversity can be explained if we consider the impact of baryonic feedback on the central dark matter densities, which depends largely on the ratio of stellar-to-halo mass as predicted by recent $N$-body and hydrodynamical simulations.
Therefore, $\Lambda$CDM framework combined with baryon physics can explain the observed dark matter densities in the classical dSphs.

We also investigate the relation between the central dark matter density profiles and their star formation histories, because several high-resolution dark matter and hydrodynamical simulations predicted the correlation between these.
However, we find no clear relation between an inner slope parameter of dark matter density profile and SFH characterized by $\tau_{0.7}$.

We confirm that a dark matter density at a radius of 150~pc is anti-correlated with the pericenter distance of a dSph suggested by \citet{2019MNRAS.490..231K}.  
Furthermore, this anti-correlations also found in the simulated dark subhalos.
In addition, we also find that the maximum circular velocities of subhalos over their formation histories, $V_{\rm peak}$ of subhalos depends slightly on $\rho_{DM}(150\ {\rm pc})$ and $r_{\rm peri}$.
This implies that the subhalos having dense central densities can survive from strong tidal effects due to being closer to the center of a host halo.

Using our non-spherical mass models, we calculate the circular velocity profiles of all sample dSphs and compare with those estimated by spherical mass models.
As a result, the shapes of circular velocity profiles, especially quantified with the maximum circular velocity, $V_{\rm max}$, are quite different between spherical and axisymmetric mass models. 
However, there are huge uncertainties in our estimated circular velocities, especially their outskirts due to the lack of data sample.

To ensure our conclusions, it is necessary to determine the dark matter density profiles for much fainter dSphs, namely ultra-faint dSphs, which are believed to have held original dark matter density profiles.
It is also important to more precisely estimate the dark matter profiles of the classical dSphs.
The next-generation wide-field spectroscopic surveys with the Subaru Prime Focus Spectrograph~\citep{2014PASJ...66R...1T} will enable us to obtain statistically significant samples of stellar kinematics and chemical abundances for the Galactic dSphs over the wide areas out to their outskirts, thereby allowing us to estimate robustly their dark matter density profiles.

\section*{Acknowledgements}
We would like to give special thanks to Manoj Kaplinghat, Ethan Nadler, Hai-Bo Yu, Chervin Laporte, Masahiro Ibe, Shigeki Matsumoto, Evan Kirby for useful discussions.
This work was supported in part by the MEXT Grant-in-Aid for Scientific Research on Innovative Areas (No.~18H04359, 18J00277 and 20H01895 for K.H., No.~17H01101, 18H04334 and 18H05437 for~M.C., No.~17H01101, 17H04828 and 18H04337 for~T.I.).
Numerical computations were partially carried out on Aterui II supercomputer at
Center for Computational Astrophysics, CfCA, of National Astronomical
Observatory of Japan.  T.I. has been supported by MEXT as
``Priority Issue on Post-K computer'' (Elucidation of the Fundamental
Laws and Evolution of the Universe), JICFuS, 
and Mext as ``Program for Promoting Researches on the Supercomputer
Fugaku'' (Toward a unified view of the universe: from large scale
structures to planets, proposal numbers hp200124).

\appendix
\restartappendixnumbering

\section{Application to mock data}\label{sec:AppA}
\subsection{Mock data}
In this section, we show the ability of our axisymmetric Jeans models to recover the dark matter density profile using mock data sets.
We focus on testing how precisely our models are able to reproduce dark matter inner slopes.

To this end, we utilize the public mock data sets provided by \cite{2016MNRAS.463.1117Z}.
They generated their mock data sets using {\sc AGAMA}\footnote{https://github.com/GalacticDynamics-Oxford/Agama} code~\citep{2019MNRAS.482.1525V}.
The data sets are generated by two kinds of dark matter halos.
One is a cusped dark halo~$(\alpha,\beta,\gamma)=(1,3,1)$ for a generalized Hernquist profile~(see equation~\ref{DMH}), while another is a cored one corresponding to $(\alpha,\beta,\gamma)=(1.5,3,0)$.
The dark matter scale densities are $\log_{10} \rho_s=-1.189$~[$M_{\odot}$~pc$^{-3}$] for cusped halo and $-0.189$ for cored one, while the scale radius in each case is equal to $\log_{10} r_s=3.0$~[pc].
For the stellar distributions, they considered axisymmetric distributions with the axial ratios, $q=0.95$ for cusped and $0.88$ for cored dark halo cases, whilst dark matter halos are spherical (e.g., $Q=1$).
The stellar motions in each halo model do not have systematic rotation, but have different velocity anisotropy parameters:
$-\log_{10}[1-\beta_z]=-0.1$ for cusped halo and $0.4$ for cored one.
Then they employed distribution functions that are expressed as double power-laws in the action integrals $\boldsymbol J$ and sampled the positions~$(x,y,z)$ and velocities~$(v_x,v_y,v_z)$ of a sample of 5000 stars for each dark matter halo model. The measurement error of velocity is fixed by 3~km~s$^{-1}$.

In order to obtain line-of-sight data, we project the system along $z$ direction with edge-on view, and place it at a distance of 80~kpc. Then we set the center of the system is at $(0,0)$~deg, and the position angle is $90^{\circ}$.
Then we randomly extract the projected positions and line-of-sight velocities of stars.

\subsection{Fitting analysis}
Before we perform MCMC analysis for mock kinematic data, we estimate stellar structural parameters such as projected axial ratio and half-light radius, which are used to calculate the Jeans equations.
To do this, we adopt the Plummer model~(see equation~\ref{plummer}) and employ a maximum likelihood algorithm~\citep{2008ApJ...684.1075M} applied to the projected positions. We do not include contamination stars for simplicity.
To perform this analysis, we set the parameters~$(\alpha_0,\delta_0,\theta_{\rm PA}, \epsilon, r_{\rm  half})$, where $(\alpha_0,\delta_0)$ are the center of the system, $\theta_{\rm PA}$ is the position angle, and $ \epsilon, r_{\rm  half}$ are the ellipticity and half-light radius, respectively.
The ellipticity is defined as $\epsilon=1-q^{\prime}$, where $q^{\prime}$ is the projected axial ratio~(see section~\ref{sec:starprof}).

The results from the fitting analysis are shown in the upper left panel of each corner plot in Figure~\ref{fig:Mock}.
We reproduce successfully $(\alpha_0,\delta_0,)=(0.0,0.0)$~deg and $\theta_{\rm PA}=90^{\circ}$.
For the ellipticity and half-light radius, we obtain $\epsilon=0.09$ and $r_{\rm half}=1100$~pc for a cusped halo, while $\epsilon=0.15$ and $r_{\rm half}=1000$~pc for a cored one.

Using the estimated stellar parameters, we perform MCMC fitting for the mock kinematic data to infer the dark matter density profiles.
For the sample size of kinematic data, we sampled 1000 and 4000 stars for comparison.
The free parameters and fitting procedure are the same as we explained in Section~\ref{sec:dmprof} and \ref{sec:fitting}.

\subsection{Model recovery}
\begin{figure}
    \centering
    \begin{minipage}{0.49\hsize}
	\begin{center}
		\includegraphics[width=\columnwidth]{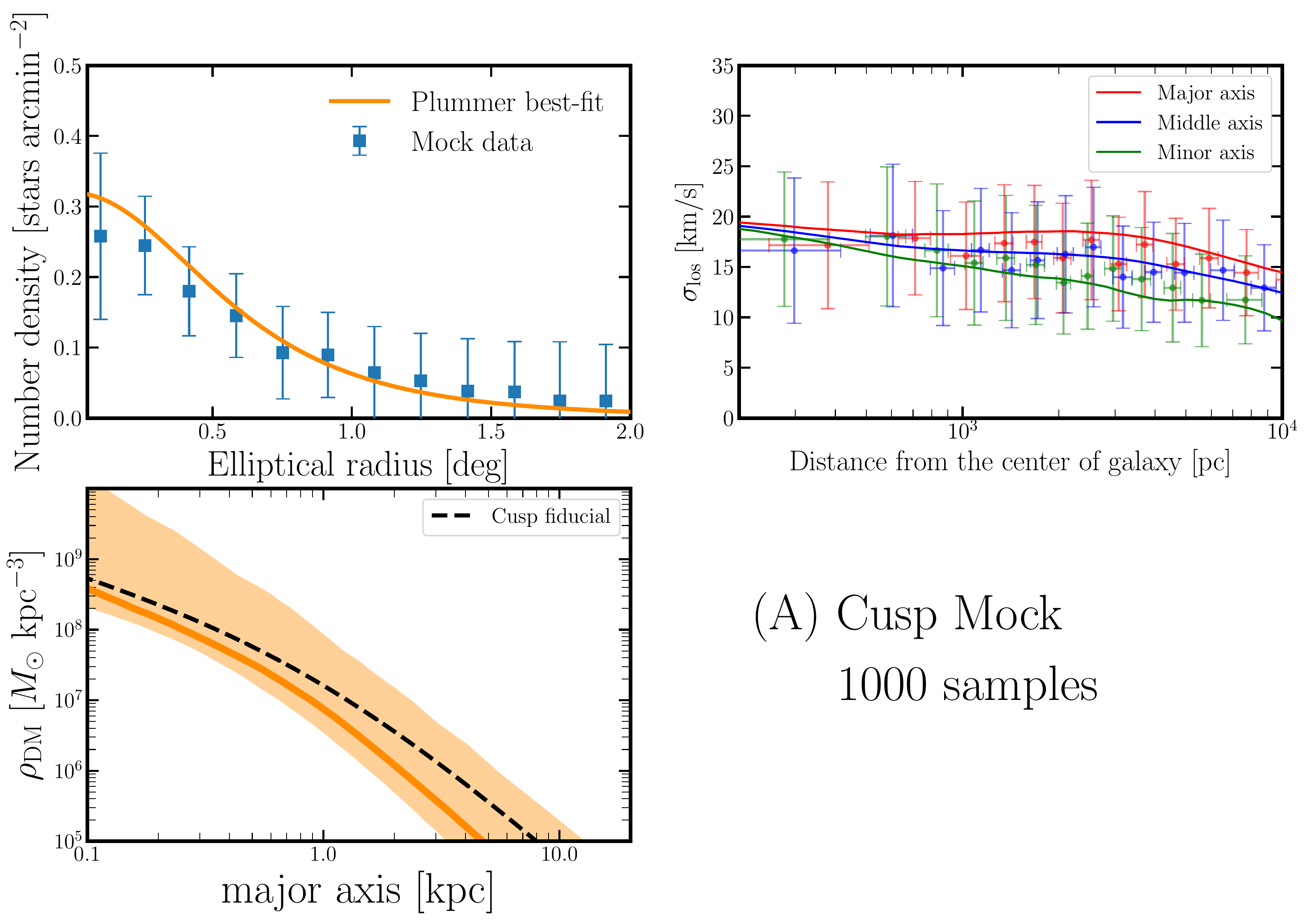}
	\end{center}
	\end{minipage}
	\begin{minipage}{0.49\hsize}
		\begin{center}
		\includegraphics[width=\columnwidth]{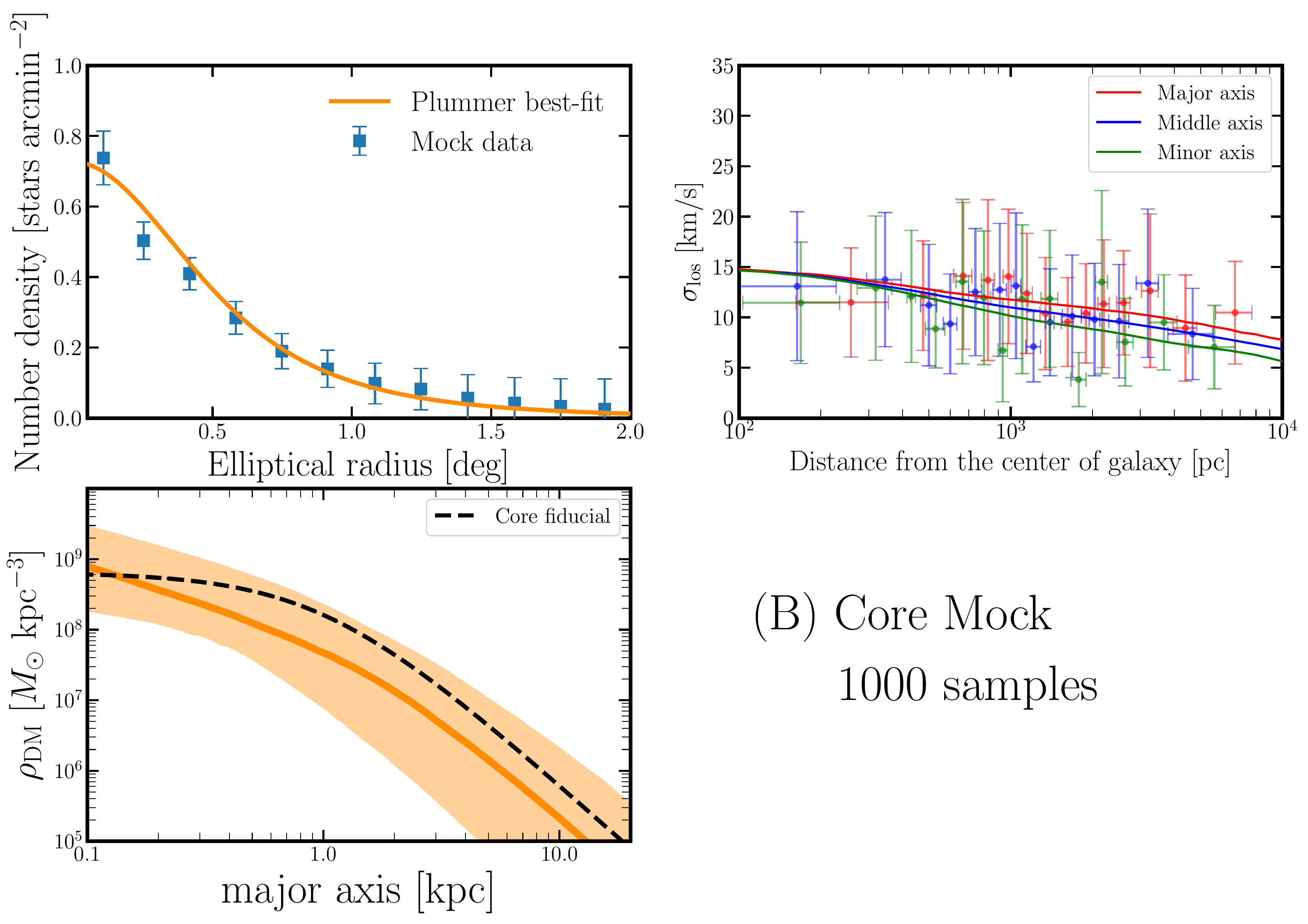}
		\end{center}
	\end{minipage}
	\begin{minipage}{0.50\hsize}
		\begin{center}
		\includegraphics[width=\columnwidth]{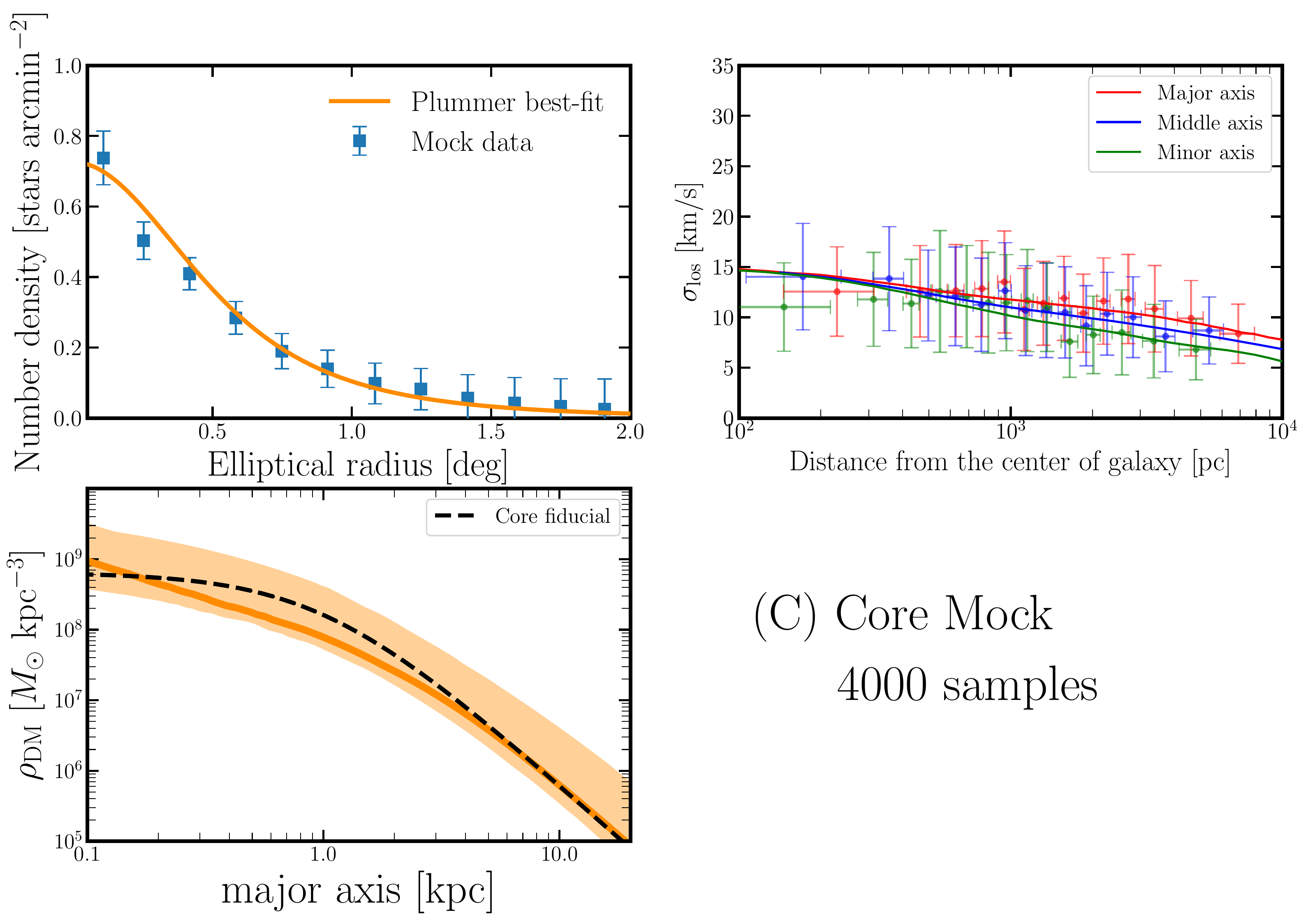}
		\end{center}
	\end{minipage}
    \caption{The results of the fitting analysis for the different mock samples: (A) a cusped dark matter halo and 1000 kinematic samples, (B) a cored one and 1000 samples, and (C) a cored one and 4000 samples.
    Each result shows the corner plots which contain the surface density profiles~(upper left), line-of-sight velocity dispersion profiles along the major, middle and minor axes of the system~(upper right), and dark matter density profiles along the major axis~(lower left), respectively.
    In the panels of the surface density profile, the solid lines are the best-fit Plummer profiles, and the points with error bars are estimated from each mock data.
    In the velocity dispersion panels, the solid lines are the median velocity dispersion of the models, while the colored points with error bars denote the mock ones.
    The orange solid lines and shaded regions on the panels depicting dark matter density profiles denote the median values and the 68 per~cent confidence intervals, in comparison with given density profiles shown with dashed lines.}
    \label{fig:Mock}
\end{figure}

Figure~\ref{fig:Mock} shows the results from the fitting analysis in the cases of (A) a cusped model with 1000 samples, (B) a cored model with 1000 samples, and (C) a cored one with 4000 samples, respectively. 
In this figure, we show the estimated line-of-sight velocity dispersion profiles and dark matter density profiles from their posterior distributions of the free parameters.
Our fitting analysis can reproduce the velocity dispersion profiles along the major, middle, and minor axes, respectively, within their $1\sigma$ errors.
The resultant dark matter density profiles are generally recovered within $1\sigma$ uncertainties for the both cases of the cusped and cored dark matter density profiles.
The estimated $\gamma$ values in each case are $\gamma=1.2^{+0.4}_{-0.5}$, $0.4^{+0.8}_{-0.3}$, and $0.4^{+0.3}_{-0.2}$ for the cusped mock~(A), the cored ones~(B) and (C), respectively.
Therefore, when we perform the fitting analysis to the cored mock data, our Jeans models indicate a somewhat biased dark matter density profile with an amount of $\Delta \gamma \simeq +0.4$, although the model is nearly consistent with a cored profile within $1\sigma$ confidence.
As for the cusped mock data with $\gamma = 1$, our models yield a smaller bias of $\Delta \gamma \simeq+0.2$ than the cored case, although, again, the model is nearly consistent with a $\gamma=1$ profile within 1$\sigma$ confidence.

\section{Posterior probability distribution functions}\label{sec:AppB}
\begin{figure*}
	\begin{minipage}{0.49\hsize}
		\begin{center}
			\includegraphics[width=\columnwidth]{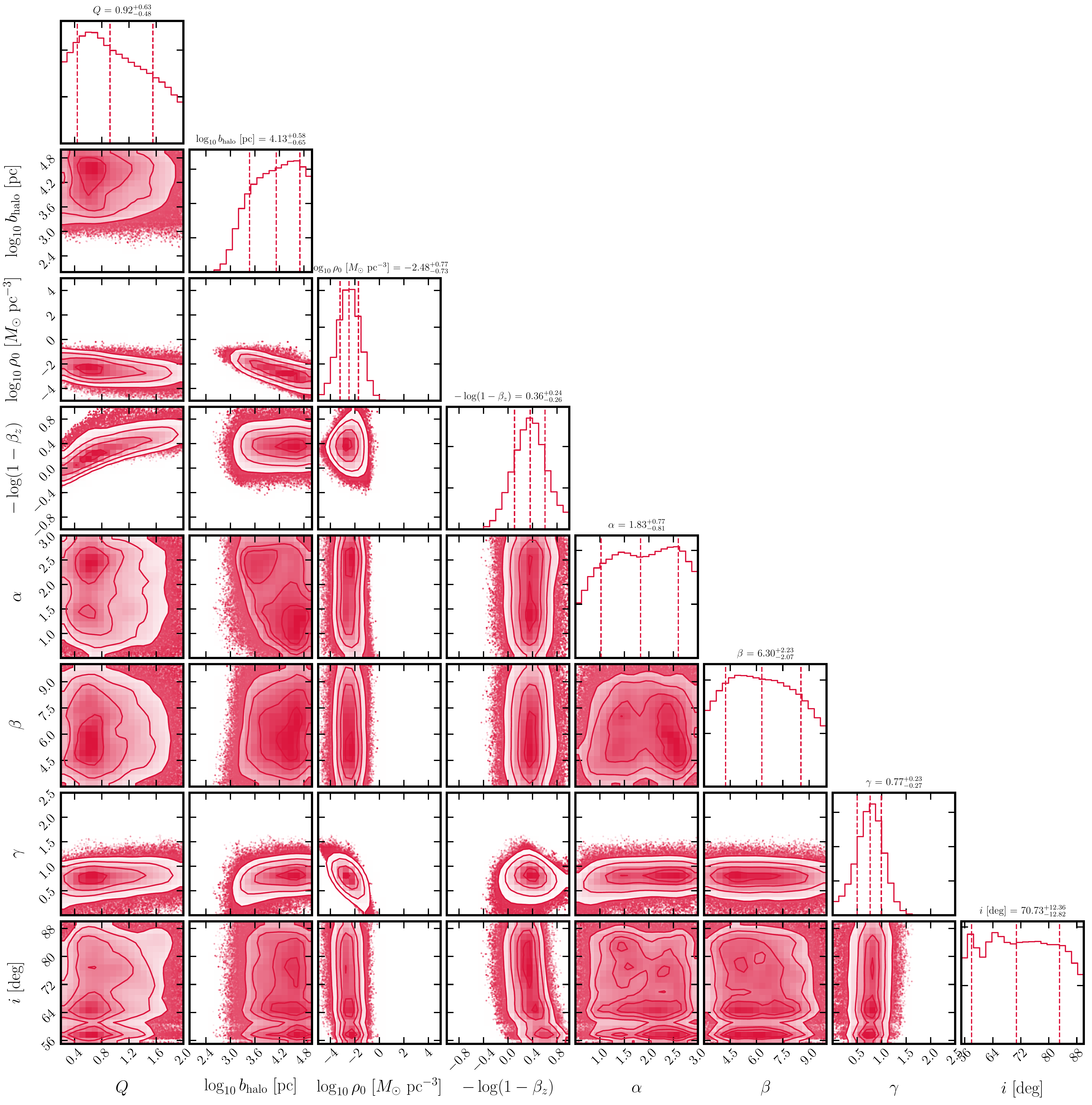}
		\end{center}
	\end{minipage}
	\begin{minipage}{0.49\hsize}
		\begin{center}
			\includegraphics[width=\columnwidth]{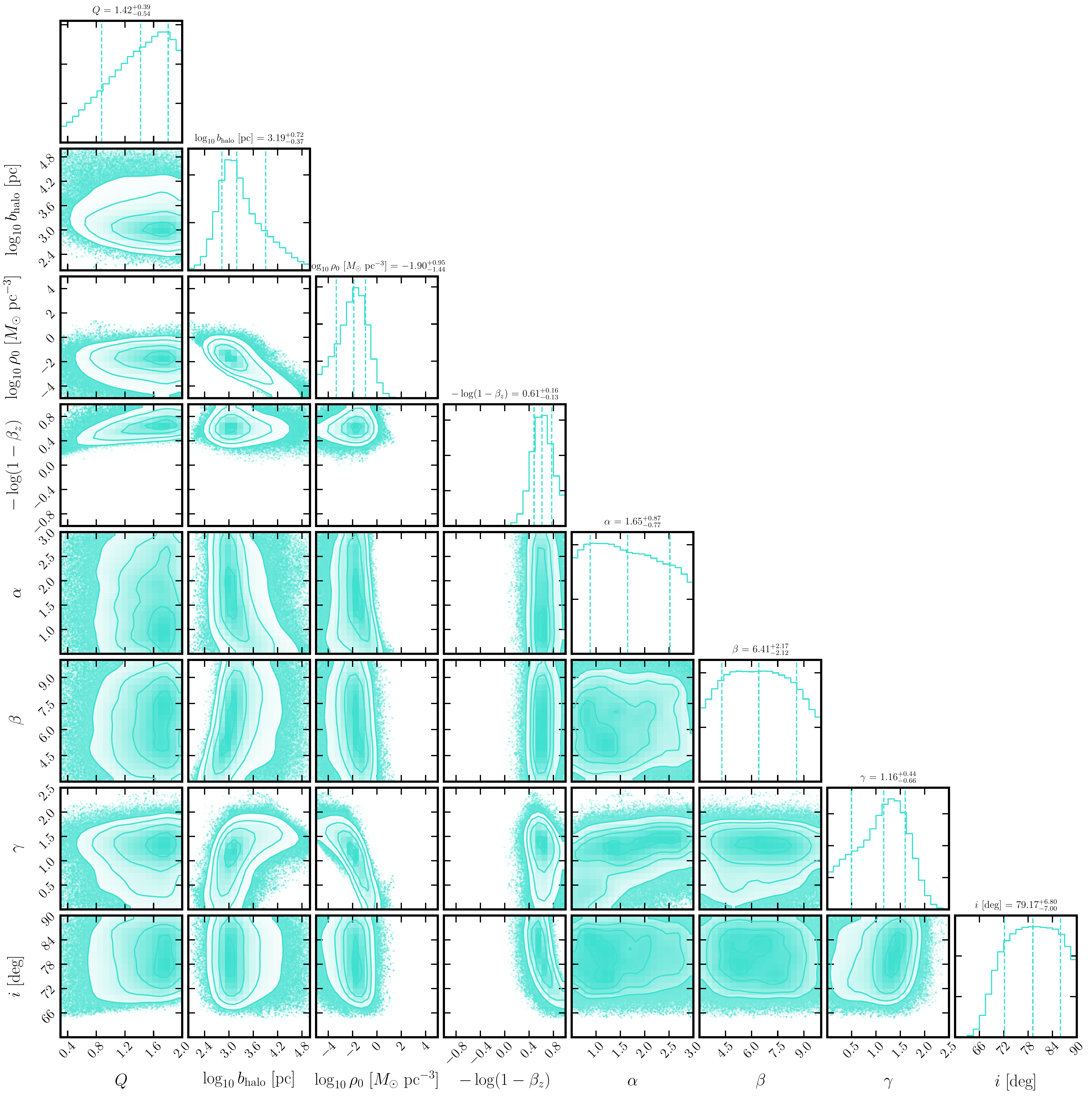}
		\end{center}
	\end{minipage}
    \caption{Same as figure~\ref{drafnx}, but for Carina~(left) and Ursa~Minor~(right).}
    \label{carumi}
\end{figure*}

\begin{figure*}
	\begin{minipage}{0.49\hsize}
		\begin{center}
			\includegraphics[width=\columnwidth]{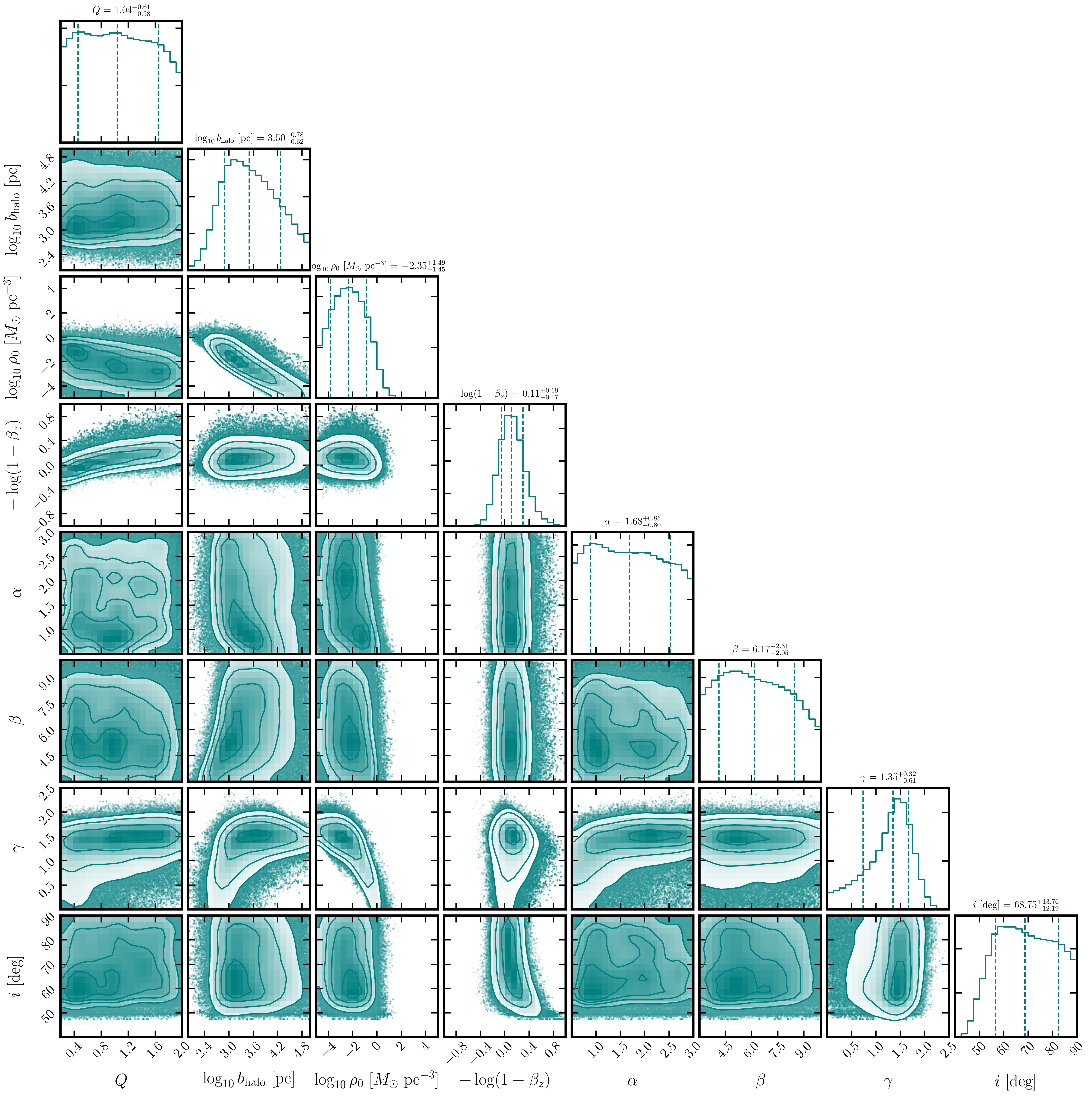}
		\end{center}
	\end{minipage}
	\begin{minipage}{0.49\hsize}
		\begin{center}
			\includegraphics[width=\columnwidth]{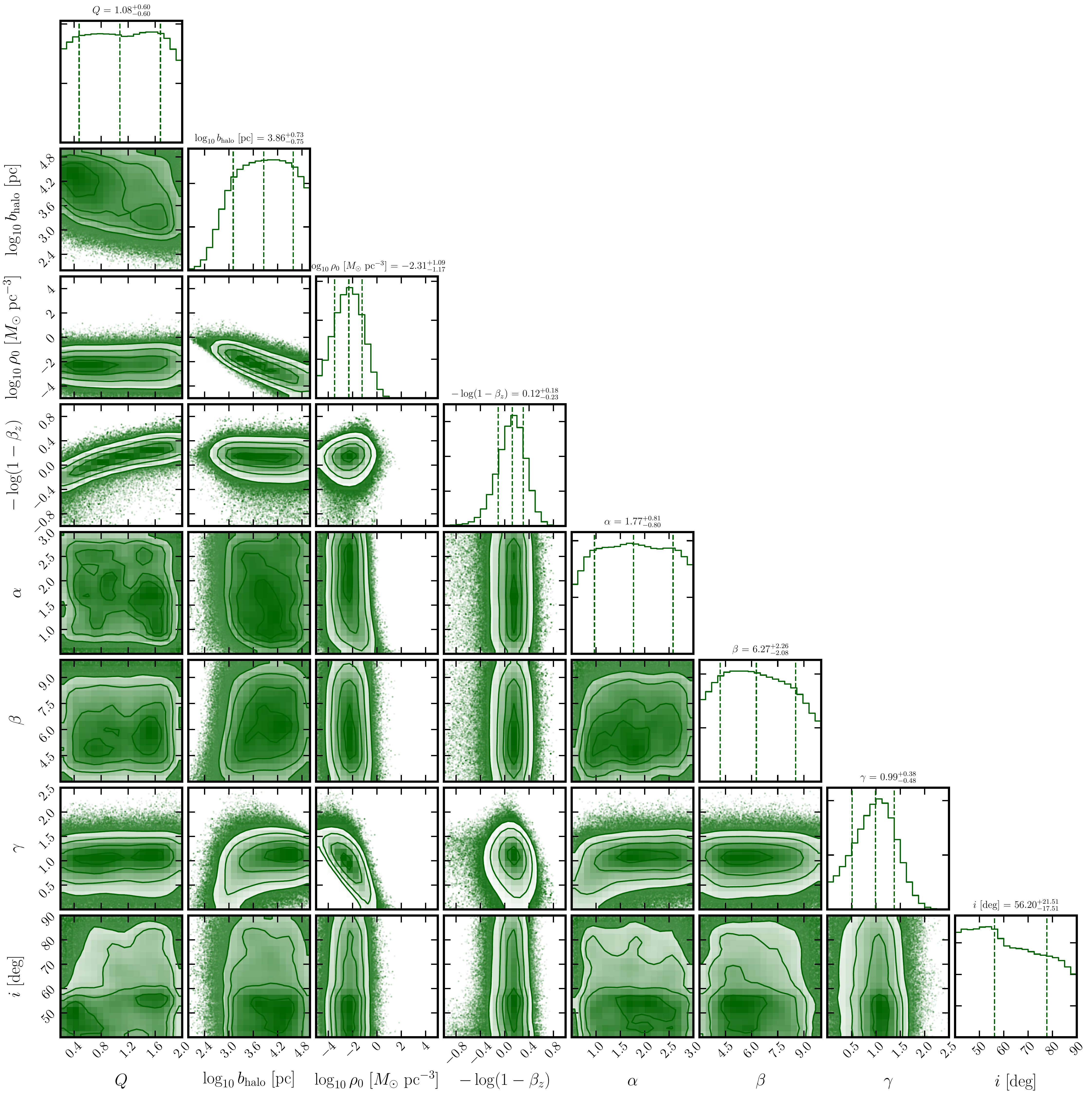}
		\end{center}
	\end{minipage}
    \caption{Same as figure~\ref{drafnx}, but for Leo I~(left) and Leo~II~(right).}
    \label{leo12}
\end{figure*}

\begin{figure*}
	\begin{minipage}{0.49\hsize}
		\begin{center}
			\includegraphics[width=\columnwidth]{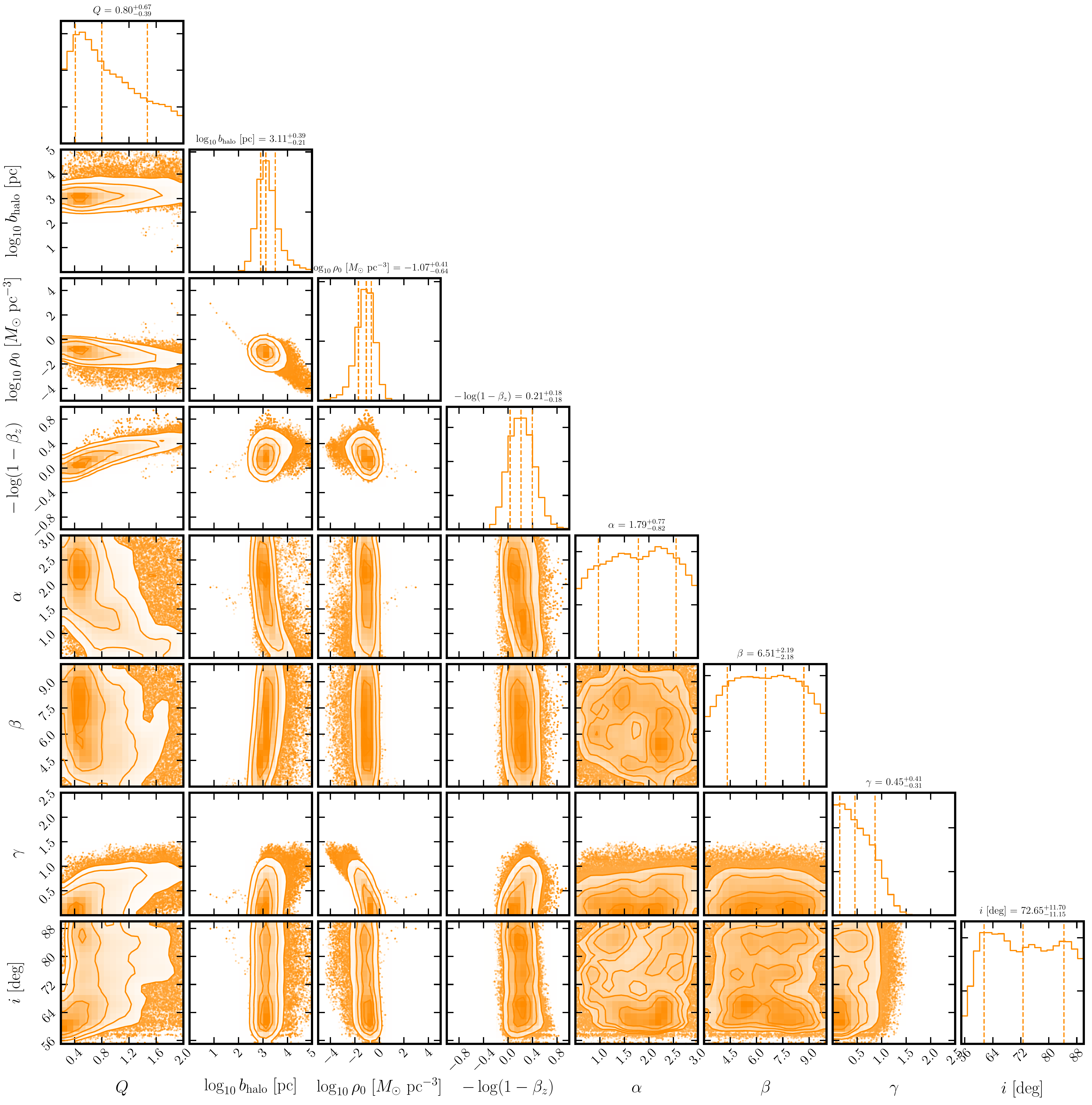}
		\end{center}
	\end{minipage}
	\begin{minipage}{0.49\hsize}
		\begin{center}
			\includegraphics[width=\columnwidth]{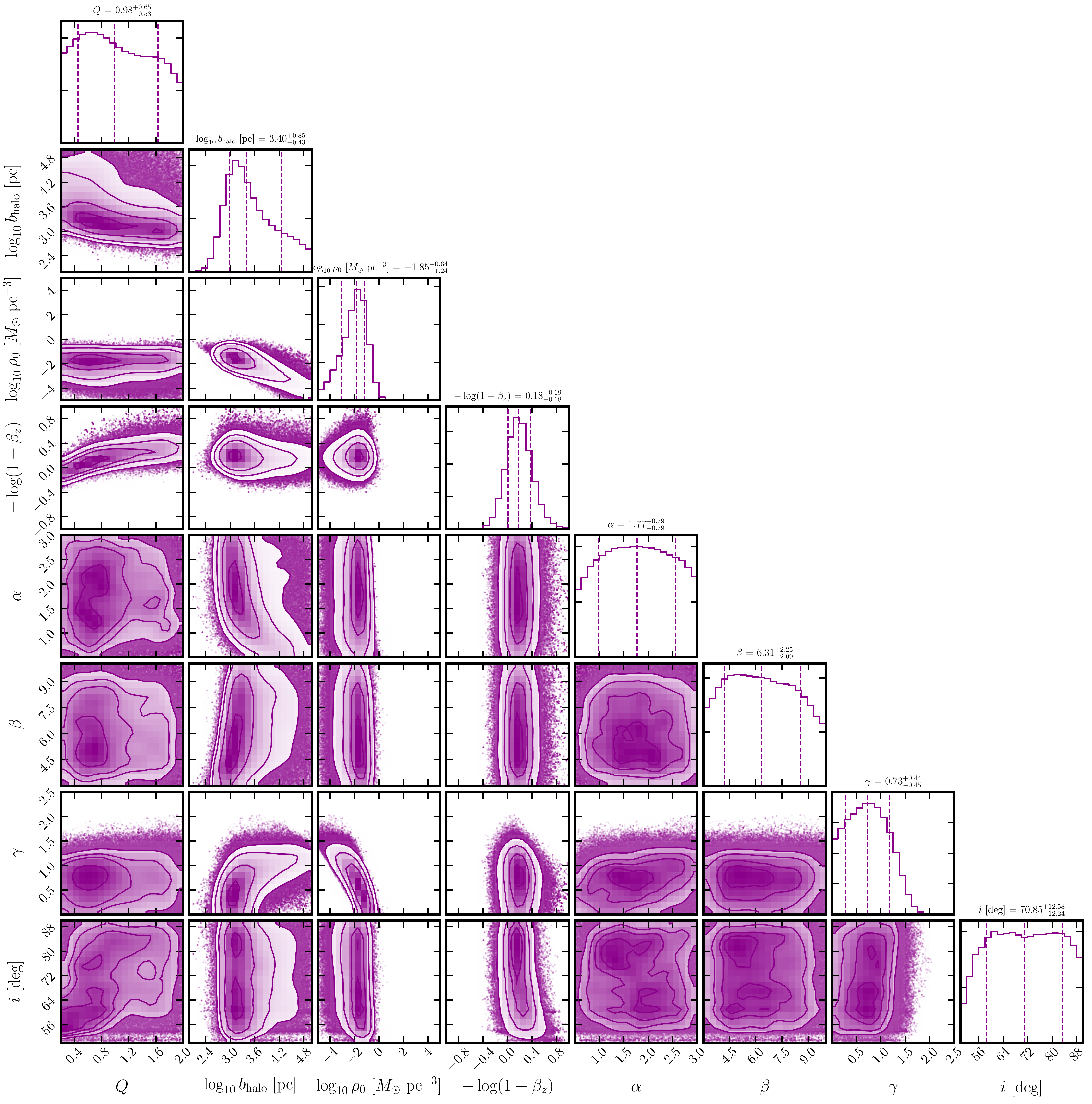}
		\end{center}
	\end{minipage}
    \caption{Same as figure~\ref{drafnx}, but for Sculptor~(left) and Sextans~(right).}
    \label{sclsex}
\end{figure*}

We here show the results of parameter estimations based on MCMC fitting analysis.
Figure~\ref{carumi}, \ref{leo12}, and \ref{sclsex} display the posterior PDFs for Carina and Ursa~Minor, for Leo~I and Leo~II, and for Sculptor and Sextans, respectively.

\section{Effects of the parameters on line-of-sight velocity dispersion profile}\label{sec:AppC}
\begin{figure*}[h!]
    \begin{center}
	\includegraphics[scale=0.5]{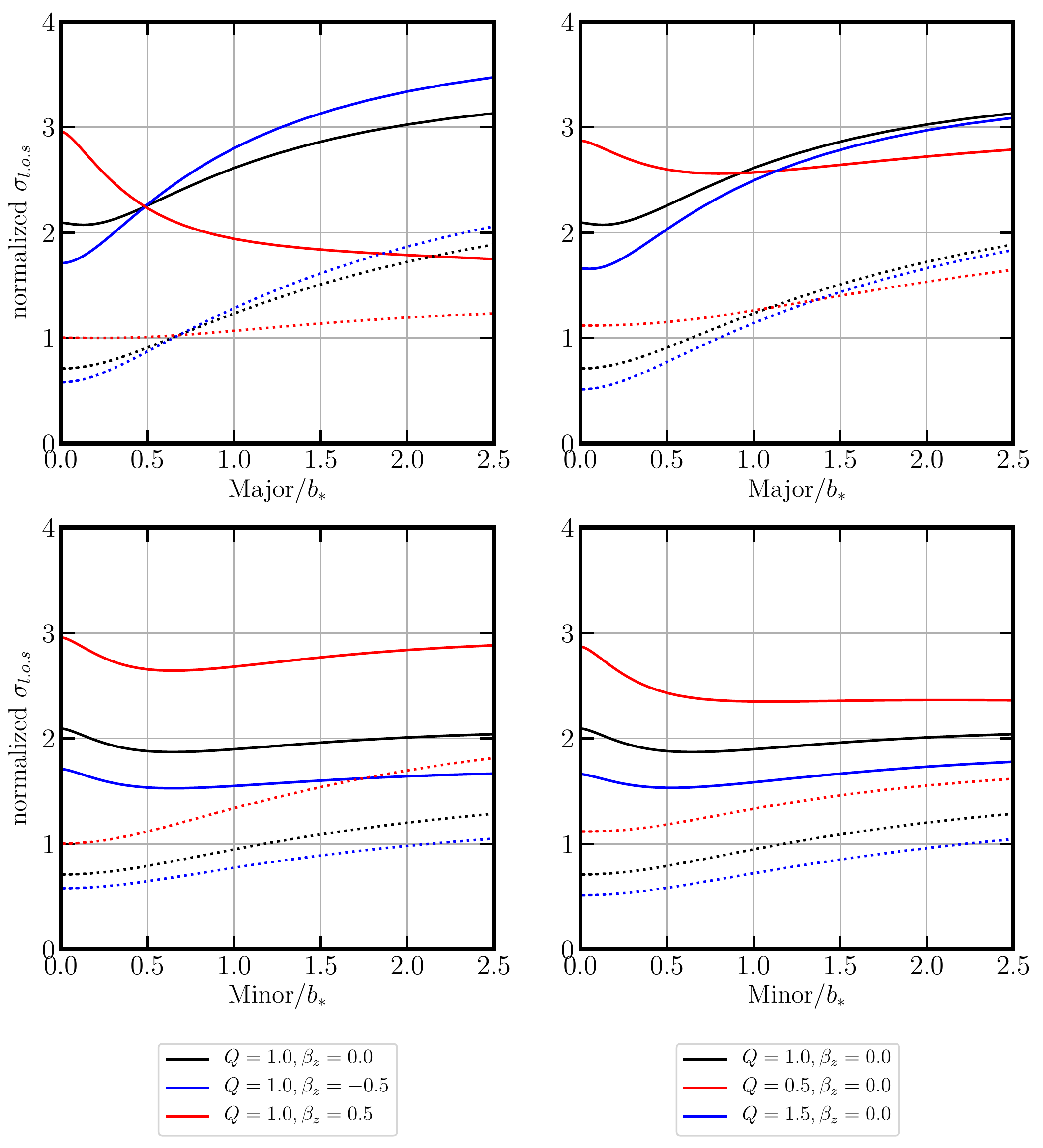}
    \caption{The upper panels show the normalized line-of-sight velocity dispersion profiles, $\sigma_{\rm l.o.s}/(Gb^2_{\ast}\rho_{s,r_s/b_{\ast}=3})^{1/2}$, along the major axis, whereas the lower panels show these profiles along the minor axis.
    The solid lines are the case for a cusped NFW dark matter density profile, while dotted lines are the case for cored ones.
    {\it Left-hand column:} the normalized $\sigma_{\rm l.o.s}$ profiles with changing $\beta_z$ under $Q=1$, 
    {\it right column:} those with changing $Q$ under $\beta_z=0$.
    For all of these cases we suppose that the oblate stellar distribution $(q=0.7)$, the edge-on galaxy $(i=90^{\circ})$, and the ratio for $r_s/b_{\ast}=3$ for the sake of demonstration.}
    \label{los_demo4}
    \end{center}
\end{figure*}

\restartappendixnumbering
Here we demonstrate the effects of the parameters~(a non-spherical shape of dark matter halo $Q$, a stellar velocity anisotropy $\beta_z$, and an inner slope of dark matter density profile, $\gamma$) on line-of-sight velocity dispersion profiles, discussed in Section~\ref{sec:demonstration}.
Figure~\ref{los_demo4} shows the normalized line-of-sight velocity dispersion profiles along the major~(top panels) and the minor axes~(bottom panels) for the oblate stellar system $(q=0.7)$, the edge-on~$(i=90^{\circ})$, and the ratio for $b_{\rm halo}/b_{\ast}=3$.
The left-hand panels show the dispersion profiles with changing the value of velocity anisotropy parameter, $\beta_z$, under spherical dark matter halo, $Q=1$, whilst the right-hand ones depict those with changing $Q$ under semi-isotropic velocity dispersion~$\beta_z=0$.
The solid lines in each panel depict the velocity dispersion profiles computed by a cusped NFW dark matter density profile, while the dotted ones are those by a cored Burkert density profile.

\bibliographystyle{aasjournal}
\bibliography{reference}{}



\end{document}

%% file: table1.tex
\begin{table*}
	\centering
	\caption{The observational data for the classical dSph galaxies.}
	\label{table1}
	\begin{tabular}{cccccccccc} 
		\hline\hline
Object & $N_{\rm sample}$ & RA(J2000) & DEC(J2000)   & $M_{\ast}$        &$D_{\odot}$ & $b_{\ast}$ &  $q^{\prime}$ & $\langle u\rangle_{obs}$ &  Ref.\\
       &                  & [hh:mm:ss] &  [dd:mm:ss] & [$10^6M_{\odot}$] &      [kpc]         &        [pc]              &    (axial ratio) & [km~s$^{-1}$]  & \\
		\hline
		Draco         & 468    & 17:20:12.4   & $+$57:54:55  & $0.29$ & $ 76\pm 6$  & $214\pm  2$  & $0.71\pm0.01$ & $-290.0$  & (1),(2),(9)\\
		Ursa~Minor    & 313   & 15:08:08.5   & $+$67:13:21   & $0.29$ & $ 76\pm 3$  & $407\pm  2$  & $0.45\pm0.01$ & $-246.9$  & (1),(3),(10)\\
		Carina        & 1086   & 06:41:36.7   & $-$50:57:58  & $0.38$ & $106\pm 6$  & $308\pm  23$  & $0.64\pm0.01$ & $220.7$  &  (1),(4),(11)\\
		Sextans       & 445    & 10:13:03.0   & $-$01:36:53  & $0.44$ & $ 86\pm 4$  & $413\pm  3$  & $0.70\pm0.01$ & 224.3  &  (1),(5),(12)\\
		Leo~I         & 328    & 10:08:28.1   & $+$12:18:23  & $5.5$ & $254\pm15$  & $270\pm  2$  & $0.70\pm0.01$ & 282.9  &  (1),(6),(13)\\
		Leo~II        & 177    & 11:13:28.8   & $+$22:09:06  & $0.74$ & $233\pm14$  & $171\pm  2$  & $0.93\pm0.01$ & $ 78.7$  &  (1),(7),(14)\\
		Sculptor      & 1360   & 01:00:09.4   & $-$33:42:33  & $2.3$ & $ 86\pm 6$  & $280\pm  1$  & $0.67\pm0.01$ & $111.4$  &  (1),(8),(12)\\
		Fornax        & 2523   & 02:39:59.3   & $-$34:26:57  & $20$ & $147\pm12$  & $838\pm  3$  & $0.71\pm0.01$ & $ 55.2$  &  (1),(4),(12)\\
	\hline
	\end{tabular}
\begin{flushleft}
References: (1)~\citet{2018ApJ...860...66M};  
(2)~\citet{2004AJ....127..861B}; (3)~\citet{2002AJ....123.3199C}; (4)~\citet{2009AJ....138..459P}; (5)~\citet{2009ApJ...703..692L}; (6)~\citet{2004MNRAS.354..708B}; (7)~\citet{2005MNRAS.360..185B}; (8)~\citet{2008AJ....135.1993P}; (9)~\citet{2015MNRAS.448.2717W}; (10)~\citet{2018AJ....156..257S}; (11)~\citet{2016ApJ...830..126F}; (12)~\citet{2009AJ....137.3100W}; (13)~\citet{2008ApJ...675..201M}; (14)~\citet{2007AJ....134..566K};
\end{flushleft}
\end{table*}

%% file: table2.tex
\begin{table*}
	\centering
	\caption{Parameter constraints for MW dSph satellites. Errors correspond to the $1\sigma$ range of our analysis.}
	\label{table2}
	\begin{tabular}{cccccccccc} 
		\hline\hline
Object  & $Q$ & $\log_{10}(b_{\rm halo})$ & $\log_{10}$($\rho_0)$   & $-\log_{10}(1-\beta_z)$ & $\alpha$  & $\beta$  & $\gamma$ & $i$   & $\rho_{\rm DM}(150 {\rm pc})$\\
        &     &            [pc]           & [$M_{\odot}$~pc$^{-3}$] &                         &           &          &          & [deg] & $10^7$[$M_{\odot}$~kpc$^{-3}$] \\
		\hline
Draco 
&$ 1.39 _{- 0.55 } ^{+ 0.40 } $ 
&$ 4.30 _{- 0.54 } ^{+ 0.46 } $ 
&$ -2.77 _{- 0.64 } ^{+ 0.64 } $
&$ 0.41 _{- 0.19 } ^{+ 0.21 } $
&$ 2.04 _{- 0.79 } ^{+ 0.64 } $
&$ 6.19 _{- 2.03 } ^{+ 2.31 } $
&$ 1.03 _{- 0.15 } ^{+ 0.14 } $
&$ 63.0 _{- 9.40 } ^{+ 16.6 } $
&$ 23.5 _{- 6.30 } ^{+ 12.8 } $\\
Ursa~Minor
&$ 1.42 _{- 0.54 } ^{+ 0.39 } $
&$ 3.19 _{- 0.37 } ^{+ 0.72 } $
&$ -1.90 _{- 1.44 } ^{+ 0.95 } $
&$ 0.61 _{- 0.13 } ^{+ 0.16 } $
&$ 1.65 _{- 0.77 } ^{+ 0.87 } $
&$ 6.41 _{- 2.12 } ^{+ 2.17 } $
&$ 1.16 _{- 0.66 } ^{+ 0.44 } $
&$ 79.2 _{- 7.01 } ^{+ 6.82 } $
&$ 23.8 _{- 7.22 } ^{+ 38.6 } $\\
Carina
&$ 0.92 _{- 0.48 } ^{+ 0.63 } $
&$ 4.13 _{- 0.65 } ^{+ 0.58 } $
&$ -2.48 _{- 0.73 } ^{+ 0.77 } $
&$ 0.36 _{- 0.26 } ^{+ 0.24 } $
&$ 1.83 _{- 0.81 } ^{+ 0.77 } $
&$ 6.31 _{- 2.07 } ^{+ 2.23 } $
&$ 0.77 _{- 0.27 } ^{+ 0.23 } $
&$ 70.7 _{- 12.8 } ^{+ 12.4 } $
&$ 10.9 _{- 3.21 } ^{+ 8.22 } $\\
Sextans
&$ 0.98 _{- 0.53 } ^{+ 0.65 } $
&$ 3.40 _{- 0.43 } ^{+ 0.85 } $
&$ -1.85 _{- 1.24 } ^{+ 0.64 } $
&$ 0.18 _{- 0.18 } ^{+ 0.19 } $
&$ 1.77 _{- 0.79 } ^{+ 0.79 } $
&$ 6.31 _{- 2.09 } ^{+ 2.25 } $
&$ 0.73 _{- 0.45 } ^{+ 0.44 } $
&$ 70.9 _{- 12.2 } ^{+ 12.6 } $
&$ 5.2 _{- 2.3 } ^{+ 3.6 } $\\
Leo~I
&$ 1.04 _{- 0.58 } ^{+ 0.61 } $
&$ 3.50 _{- 0.62 } ^{+ 0.78 } $
&$ -2.35 _{- 1.45 } ^{+ 1.49 } $
&$ 0.11 _{- 0.17 } ^{+ 0.19 } $
&$ 1.68 _{- 0.80 } ^{+ 0.85 } $
&$ 6.17 _{- 2.05 } ^{+ 2.31 } $
&$ 1.35 _{- 0.61 } ^{+ 0.32 } $
&$ 68.7 _{- 12.2 } ^{+ 13.8 } $
&$ 26.4 _{- 9.10 } ^{+ 22.3 } $\\
Leo~II
&$ 1.08 _{- 0.60 } ^{+ 0.61 } $
&$ 3.86 _{- 0.75 } ^{+ 0.73 } $
&$ -2.31 _{- 1.17 } ^{+ 1.09 } $
&$ 0.12 _{- 0.23 } ^{+ 0.18 } $
&$ 1.77 _{- 0.81 } ^{+ 0.81 } $
&$ 6.27 _{- 2.08 } ^{+ 2.26 } $
&$ 0.99 _{- 0.48 } ^{+ 0.38 } $
&$ 56.2 _{- 17.5 } ^{+ 21.5 } $
&$ 20.2 _{- 6.10 } ^{+ 12.7 } $\\
Sculptor
&$ 0.82 _{- 0.39 } ^{+ 0.67 } $
&$ 3.11 _{- 0.21 } ^{+ 0.39 } $
&$ -1.07 _{- 0.64 } ^{+ 0.41 } $
&$ 0.21 _{- 0.18 } ^{+ 0.18 } $
&$ 1.79 _{- 0.82 } ^{+ 0.77 } $
&$ 6.51 _{- 2.18 } ^{+ 2.19 } $
&$ 0.45 _{- 0.31 } ^{+ 0.41 } $
&$ 72.6 _{- 11.2 } ^{+ 11.7 } $
&$ 21.4 _{- 6.30 } ^{+ 12.6 } $\\
Fornax
&$ 1.04 _{- 0.58 } ^{+ 0.63 } $
&$ 3.27 _{- 0.21 } ^{+ 0.41 } $
&$ -1.54 _{- 0.47 } ^{+ 0.31 } $
&$ 0.24 _{- 0.18 } ^{+ 0.13 } $
&$ 1.98 _{- 0.74 } ^{+ 0.64 } $
&$ 6.63 _{- 2.15 } ^{+ 2.11 } $
&$ 0.44 _{- 0.29 } ^{+ 0.40 } $
&$ 72.3 _{- 11.6 } ^{+ 11.5 } $
&$ 12.2 _{- 2.30 } ^{+ 3.24 } $\\
	\hline
	\end{tabular}
\end{table*}

%% file: table3.tex
\begin{table}[t!!]
	\centering
	\caption{The observational data for the classical dSph galaxies.}
	\label{table3}
	\begin{tabular}{ccc} 
		\hline\hline
        Object & $\log_{10}[J_{0.5}]$  & $\log_{10}[D_{0.5}]$\\
               & [GeV$^{2}$~cm$^{-5}$] & [GeV~cm$^{-2}$]\\
		\hline
	Draco         & $19.03^{+0.37}_{-0.28}$ &$18.86^{+0.21}_{-0.19}$ \\
	Ursa~Minor    & $18.59^{+0.44}_{-0.26}$ &$18.09^{+0.17}_{-0.12}$ \\
	Carina        & $18.23^{+0.43}_{-0.33}$ &$18.49^{+0.29}_{-0.27}$ \\
	Sextans       & $18.01^{+0.31}_{-0.20}$ &$18.16^{+0.24}_{-0.20}$ \\
	Leo~I         & $17.16^{+0.51}_{-0.47}$ &$17.60^{+0.48}_{-0.49}$ \\	Leo~II        & $17.75^{+0.66}_{-0.59}$ &$17.32^{+0.53}_{-0.62}$ \\
	Sculptor      & $18.54^{+0.24}_{-0.20}$ &$18.31^{+0.14}_{-0.12}$ \\
	Fornax        & $17.91^{+0.22}_{-0.13}$ &$18.11^{+0.19}_{-0.13}$ \\
	\hline
	\end{tabular}
\end{table}